\documentclass[a4paper, twocolumn, accepted=2024-11-07, linenumbers, nopdfoutputerror]{quantumarticle}
\usepackage{graphicx, amsfonts, amsmath, amssymb, physics, bm, amsthm}
\usepackage{xcolor}
\usepackage[hidelinks]{hyperref}
\usepackage{cleveref}
\usepackage{quantikz}
\usepackage{subfigure}

\newtheorem{theorem}{Theorem}[section]
\newtheorem{lemma}{Lemma}[section]
\newtheorem{definition}{Definition}[section]
\newtheorem{corollary}{Corollary}[section]

\newcommand{\btheta}{\ensuremath{{\bm\theta}}}

\newcommand{\Btheta}{\ensuremath{{\bm\Theta}}}
\newcommand{\bigO}[1]{\ensuremath{\mathcal{O}\left( #1 \right)}}

\newcommand{\Var}[2][]{\ensuremath{\operatorname{Var}_{#1}\left( #2 \right)}}
\newcommand{\Cov}[2][]{\ensuremath{\operatorname{Cov}_{#1}\left( #2 \right)}}
\newcommand{\E}[2][]{\ensuremath{\mathbb E_{#1}\left( #2 \right)}}
\newcommand{\Er}[1]{\ensuremath{\mathbb E_{\Btheta_{R, j}}\left( #1 \right)}}
\newcommand{\Erl}[1]{\ensuremath{\mathbb E_{\Btheta_{R, l}}\left( #1 \right)}}

\newcommand{\El}[1]{\ensuremath{\mathbb E_{\Btheta_{L, j}}\left( #1 \right)}}
\newcommand{\Ex}[1]{\ensuremath{\mathbb E_{X}\left( #1 \right)}}
\renewcommand{\L}{\mathcal L}
\newcommand{\B}{\mathcal B}
\newcommand{\A}{\mathcal A}
\newcommand{\D}{\mathcal D}
\newcommand{\Dir}{\operatorname{Dir}}
\newcommand{\N}{\mathcal N}
\newcommand{\K}{\mathcal K}
\newcommand{\SU}{\mathcal{SU} }

\renewcommand{\max}{\ensuremath{{\rm max}}}
\renewcommand{\min}{\ensuremath{{\rm min}}}
\newcommand{\Dirichlet}[1][\bm\alpha]{\ensuremath{{\rm Dir}\left(#1\right)}}

\newcommand{\aqa}{$\langle aQa ^L\rangle $ Applied Quantum Algorithms, Universiteit Leiden}
\newcommand{\lorentz}{Instituut-Lorentz, Universiteit Leiden, the Netherlands}
\newcommand{\cern}{Quantum Technology Initiative, CERN, Geneva, Switzerland}

\begin{document}

\title{Gradients and frequency profiles of quantum re-uploading models}

\author{Alice Barthe}
\affiliation{\aqa}
\affiliation{\cern}
\affiliation{\lorentz}
\author{Adrián Pérez-Salinas}
\affiliation{\aqa}
\affiliation{\lorentz}

\begin{abstract}
Quantum re-uploading models have been extensively investigated as a form of machine learning within the context of variational quantum algorithms. 
Their trainability and expressivity are not yet fully understood and are critical to their performance.
In this work, we address trainability through the lens of the magnitude of the gradients of the cost function. 
We prove bounds for the differences between gradients of the better-studied data-less parameterized quantum circuits and re-uploading models. 
We coin the concept of {\sl absorption witness} to quantify such difference.
For the expressivity, we prove that quantum re-uploading models output functions with vanishing high-frequency components and upper-bounded derivatives with respect to data. 
As a consequence, such functions present limited sensitivity to fine details and offer protection against overfitting. 
We performed numerical experiments extending the theoretical results to more relaxed and realistic conditions. 
Overall, future designs of quantum re-uploading models will benefit from the strengthened knowledge delivered by the uncovering of absorption witnesses and vanishing high frequencies. 
\end{abstract}
\maketitle

\section{Introduction}

Variational Quantum Algorithms (VQAs) have emerged as a prominent paradigm in the realm of quantum computing as a hybrid computational model suited for NISQ (Noisy Intermediate-Scale Quantum)~\cite{preskill2018quantum} devices in conjunction with classical optimization techniques~\cite{bharti2022noisy, cerezo2021variational}. These algorithms rely on the minimization of cost functions~\cite{mcclean2021low, bittel2021training}, which encode specific computational problems. VQAs have been used to solve a variety of problems, including approximating ground states~\cite{peruzzo2014variational, mcclean2016theory, ryabinkin2019constrained}, combinatorial challenges~\cite{farhi2014quantum}, chemistry problems~\cite{cao2019quantum} and simulation of quantum systems~\cite{li2017efficient, cirstoiu2020variational, bharti2021quantumassisted, mcardle2019variational, yuan2019theory}. Furthermore, VQAs have served as quantum computing engines for tackling various machine learning (ML) tasks, such as function regression~\cite{mitarai2018quantum, schuld2020circuitcentric}, classification~\cite{havlicek2019supervised, schuld2021supervised, otterbach2017unsupervised} or generative models~\cite{zoufal2019quantum, zoufal2021variational, dallaire-demers2018quantum}. We specifically make a distinction between linear models on the one side, introducing data either as input states or through encoding maps~\cite{schuld2019quantum}, and quantum re-uploading (QRU) schemes on the other side, which introduce data iteratively throughout the execution of the quantum circuit~\cite{vidal2020input, schuld2021effect, perez-salinas2020data, perez-salinas2021determining}. 

The performance of VQAs hinges on two critical properties: expressivity and trainability. Expressivity embodies the model's ability to represent precise solutions to the underlying problem, while trainability is a measure of the difficulty in finding the parameter set that yields the optimal attainable solution within the model. In the case of data-independent VQAs, expressivity can be intuitively understood as the proportion of attainable output states within the Hilbert space, quantified through closeness to $t$-designs~\cite{sim2019expressibility}. In the context of data-dependent ML, expressivity pertains to the suitability of the output function in fitting the data~\cite{cybenko1989approximation, hornik1991approximation}. The universality of QRU models has been proven even with a single qubit~\cite{schuld2021effect, perez-salinas2021one}. Trainability in VQAs, on the other hand, is closely linked to characteristics of the cost function, such as non-convexity~\cite{anschuetz2022barren} or vanishing gradients~\cite{mcclean2018barren}. The relationship between the trainability of VQAs and QML schemes has been previously explored, in the absence of re-uploading~\cite{thanasilp2023subtleties}. Importantly, trainability and expressivity are usually mutually exclusive, and for VQAs in particular there exists a well-studied trade-off between these two properties~\cite{holmes2022connecting,hubregtsen2021evaluation}. 

In this work, we specifically explore QRU models with a focus on exploring trainability and expressivity. 
Our investigation into trainability focuses on the on-average behavior of gradients, which can be related to the flatness of the cost function. 
We compare the cost functions of QRU models and base PQCs, which are circuits with the same architecture and observable as the QRU, but where the data gates are removed.
The difference between the flatness of both cost functions is upper bounded by a quantity we refer to as \textit{absorption witness}, which quantifies the influence of data gates on the quantum circuit when averaged over the parameter space.
Such derivation opens a path to transfer existing knowledge about the flatness of PQCs~\cite{mcclean2018barren, larocca2022diagnosing, larocca2021theory} to guide the design of QRU models.

The second segment of our findings is related to the expressivity of data-dependent output functions generated by QRU models. It is known that any hypothesis function output by QRU models can be expressed as a generalized trigonometric polynomial, with the range of available frequencies contingent on the data encoding scheme~\cite{caro2021encodingdependent, schuld2021effect}. We show that, under reasonable assumptions, the average magnitude of individual frequency components in the hypothesis function rapidly tends to a Gaussian profile, with a variance scaling as $\sim \sqrt L$, with $L$ being the number of re-uploading steps, while the support in frequencies scales as $\sim L$.  This property inherently biases the attainable hypothesis functions as being heavily dominated by lower-frequency components. This has a direct consequence on the Lipschitz constants of these output functions.

The paper is organized as follows. ~\Cref{sec.background} introduces relevant concepts and notation for the paper.~\Cref{sec.gradients} delves into the expected norm of gradients in QRU models.~\Cref{sec.expressivity} delves into the expressive capabilities of output functions within QRU models in terms of spectrum.Both sections are supported by numerical experiments showcasing agreement with our theoretical findings.
~\Cref{sec.discussion} engages in a discussion of the implications and potential avenues opened up by our research. Conclusions are summarized in~\Cref{sec.conclusions}.

\section{Background}\label{sec.background}

In this work, we refer to a PQC as a sequence of parameterized gates and fixed gates applied to an initial state, namely
\begin{equation}
    U(\btheta) = \prod_{j = 1}^M  \ W_j e^{i V_j\theta_i}, 
\end{equation}
where $\{V_j\}$ are, without loss of generality, traceless Hermitian matrices known as generators, $\{W_j\}_{j = 1}^M$ are fixed unitary operations, and $\btheta\in\Btheta \subset \mathbb R^M$. We use these PQCs as a baseline for QRU models~\cite{perez-salinas2020data, perez-salinas2021one, schuld2021effect}. This model consists of a PQC where data-encoding gates have been added in the form 
\begin{equation}\label{eq.ru_model}
        U(\btheta, x) = \prod_{j = 1}^M  \ e^{i g_j x} \ W_j e^{i V_j\theta_i}, 
    \end{equation}
where $\{g_j\}_{j = 1}^M$ are, without loss of generality, traceless Hermitian generators. We do not impose constraints in $\{V_j \} \cap \{g_j\}$. The input $x$ is a real number. Extensions to multidimensional values of $x$ are available by adding extra terms to the model, although this case will not be considered in this work. Notice that there exists a mapping between PQCs with data as initial state and QRU models, thus making both computations formally equivalent~\cite{jerbi2023quantum}, up to overheads. 

QRU models yield $\btheta$-dependent hypothesis functions 
\begin{equation}\label{eq.hypothesis_function}
        h_\btheta(x) = \bra{0} U^\dagger(\btheta, x)H U(\btheta, x) \ket{0},
\end{equation}
when applied to an initial quantum state and measured with an observable $H$. Notice that $h_\btheta(x=x_0)$ for a fixed value $x_0$ is the standard definition of the cost function of a PQC. In the case $x_0 \neq 0$, we can recover our formulation of PQC by adapting the fixed gates $W_j$. The values of $\btheta$ are trainable to match data coming in pairs $X = \{(x, y(x)\}$, such that $h_\btheta(x) \approx y(x)$. 
These hypothesis functions can be expressed as a generalized trigonometric polynomial ~\cite{schuld2021effect, caro2021encodingdependent}, namely
\begin{equation}\label{eq.hyp_function_fourier}
    h_\btheta(x) = \sum_{\omega \in \Omega} a_{\omega}(\btheta) e^{i \omega x},
\end{equation}
where $a_{\omega}(\btheta) = a_{-\omega}^*(\btheta)$ to ensure real valued hypothesis functions, and $\Omega$ is the set of available frequencies.

The training of VQAs involves an optimization procedure where a parameter set minimizing a cost function is searched. The difficulty of this optimization task is enclosed under the broad concept of trainability. A paradigmatic example of optimizing a PQC is minimizing $h_\btheta(x = 0)$ with respect to $\btheta$ to find an approximation to the ground state of the corresponding Hamiltonian $H$. Trainability has been extensively studied in the context of PQCs~\cite{anschuetz2022barren, mcclean2018barren}. Training a QRU model involves finding the optimal set of parameters $\btheta$ for which $h_\btheta(x)$ approximately matches some target function given by data. Trainability may depend on several features of the cost function landscape~\cite{munoz2015exploratory}, such as small gradients~\cite{pascanu2013difficulty} or non-convexity, e.g. the existence of (many) local minima~\cite{friedrich2022escaping}. In this work, we focus on average behaviors of gradients of the cost function, inspired by the well-studied phenomenon of vanishing gradients barren plateaus (BP)~\cite{mcclean2018barren, cerezo2021variational, holmes2022connecting}. 

Expressivity is another crucial aspect of parameterized models, capturing their ability to represent various solutions. For PQCs, expressivity entails the existence of parameter sets $\btheta^*$ making $U(\btheta^*, 0)$ close to some unitary operations $V \in \SU(2^n)$, within a specified tolerance and respect to some distance. Expressivity is often measured relative to unitary $t$-designs~\cite{sim2019expressibility}. In contrast, expressivity in the context of ML (e.g. QRU models) is related to the output function and its capability. A model is expressive if its output is able to match a variety of target functions to fit some data~\cite{raghu2017expressive}.

\section{Gradients in QRU models}\label{sec.gradients}

{\subsection{Losses for PQC and QRU}
In this section, we focus on characterizing gradients of QRU models, as compared to those of PQCs. 
In the case of PQC, the gradients of interest are typically defined in relationship to their cost function $h_\btheta(0)$.
However, the optimization of QRU models involves a cost function that depends on both the quantum circuit, expressed through $h_\btheta(x)$, and the available data, provided in pairs as $(x, y(x))$. Such cost function is usually given by averaging a distance between functions $\Delta(\cdot, \cdot)$ as
\begin{equation}\label{eq.loss_function}
    \L_X(\btheta) = \Ex{\Delta(h_\btheta(x), y(x))},
\end{equation}
where $\Ex{\cdot}$ denotes expectation value over the training dataset $X = \{(x, y(x))\}$, usually composed by a discrete set of points.
Notice that $\L_X$ is empirical as the training dataset is drawn from an unknown data distribution $X \sim \D$, and approximates the true unaccessible risk averaged over $\D$. 
In regression tasks, a common choice for the distance metric $\Delta(\cdot, \cdot)$  is the mean squared error.

Our interest lies in examining the gradients of the loss function of QRU models, expressed as
\begin{equation}\label{eq.der_loss}
    \partial_j \L(\btheta) = \Ex{ \frac{\partial \Delta(h_\btheta(x), y(x))}{\partial h_\btheta(x)} \partial_j h_\btheta(x)}.
\end{equation}
The influence imposed by the choice of distance function $\Delta(\cdot, \cdot)$ can be readily bounded e. g. using its Lipschitz constant $L_\Delta$. In particular, 
\begin{equation}\label{eq.bound_der_loss}
    \Var[\Btheta]{\partial_j \L(\btheta)} \leq L_\Delta^2 \Var[\Btheta]{\Ex{\partial_j h_\btheta(x)}}.
\end{equation}
Therefore, we can bound vanishing gradients by studying only $\Var[\Btheta]{\Ex{\partial_j h_\btheta(x)}}$. It is important to highlight that all results presented in this section are applicable for any distribution over parameters $\Btheta$.

\subsection{Gradient of the loss function}
We connect now the gradients of loss functions for QRU models and PQCs. First, the average of derivatives of hypothesis functions are zero, namely~\cite{holmes2022connecting}
\begin{equation}
    \E[\Btheta]{\Ex{\partial_j h_\btheta(x)}} = \Ex{\E[\Btheta]{\partial_j h_\btheta(x)}} = 0,
\end{equation}
if the parameters $\btheta$ are sampled uniformly from $\Btheta$. As a consequence, due to the convexity of the square function, we have $\E{x}^2 \leq \E{x^2}$. By combining these two observations and the definition $\Var{x} = \E{x^2} - \E{x}^2$, we derive
\begin{equation}\label{eq.var_x}
    \Var[\Btheta]{\Ex{\partial_j h_\btheta(x)}} \leq \Ex{\Var[\Btheta]{\partial_j h_\btheta(x)}},
\end{equation}
which can be readily connected to $\Var[\Btheta]{\partial_j\L(\btheta)}$ via~\Cref{eq.bound_der_loss}. Therefore, we can bound the variance of cost functions in QRU by the average of variances cost functions of several PQC, defined by different fixed $x_0$. Bounds on $\Var[\Btheta]{\partial_j h_\btheta(0)}$ have been studied in the context of PQC. In particular, BPs are defined for exponentially vanishing bounds to $\Var[\Btheta]{\partial_j h_\btheta(0)}$~\cite{mcclean2018barren}. 

\Cref{eq.var_x} suggests that QRU models present vanishing gradients if the base PQC presents BPs, which means that it is recommendable to use architectures that avoid vanishing gradients such as in~\cite{larocca2021theory, schatzki2022theoretical} when designing QRUs. This statement is made from a purely trainability point of view, as it has been shown that such architectures are indeed classically simulable. We are only highlighting that given a QRU model, if removing the reuploading gates yields a PQC architecture known to suffer from vanishing gradients, then vanishing gradients will also affect the QRU architecture. In addition,~\Cref{eq.var_x} does not guarantee non-vanishing gradients for QRU models derived from BP-free PQCs. As an example, consider a re-uploading model with parameterized single-qubit gates and data-encoding entangling gates arranged in an alternate-layered structure, measured by a sum of 1-local observables, as illustrated in~\Cref{fig.circuits}, base PQC 1. A compatible base PQC is composed only of single-qubit parameterized gates. In this case, $h_\btheta(x_0)$ from the base PQC has large gradients~\cite{cerezo2021cost}. The inclusion of data increases the accessible Hilbert space due to the presence of entangling operations, entanglement, which causes BPs~\cite{holmes2022connecting}.

\begin{figure}
    \centering
    ~ Re-uploading ~ \\ 
    \begin{quantikz}[row sep=0.1cm, column sep=0.2cm]
        \lstick{$\ket{0}$} & \qw & \gate{U} & \ctrl{1} & \gate{U} & \qw & \gate{X R_z(x)} & \qw & \qw & \rstick{$\cdots$} \\
        \lstick{$\ket{0}$} & \qw & \gate{U} & \gate{X R_z(x)} & \gate{U} & \ctrl{2} & \qw & \qw & \qw & \rstick{$\cdots$} \\[.2cm]
        \lstick{$ \vdots $} & \wave &&&&&& \\ [.25cm]
        \lstick{$\ket{0}$} & \qw & \gate{U} & \ctrl{1} & \gate{U} & \gate{X R_z(x)} & \qw & \qw & \qw & \rstick{$\cdots$} \\
        \lstick{$\ket{0}$} & \qw & \gate{U} & \gate{X R_z(x)} & \gate{U} & \qw & \ctrl{-4} & \qw & \qw & \rstick{$\cdots$}
    \end{quantikz}
    \vspace{5mm} \\
    ~\hspace{1cm} Base PQC 1 \hspace{2cm} Base PQC 2 \hspace{1.5cm} ~ \\ 
    \begin{quantikz}[row sep=0.1cm, column sep=0.1cm]
        \lstick{$\ket{0}$} & \qw & \gate{U} & \qw & \gate{U} & \qw & \qw& \qw & \qw & \rstick{$\cdots$} \\
        \lstick{$\ket{0}$} & \qw & \gate{U} & \qw & \gate{U} & \qw & \qw & \qw & \qw & \rstick{$\cdots$} \\[.2cm]
        \lstick{$ \vdots $} & \wave &&&&&& \\ [.25cm]
        \lstick{$\ket{0}$} & \qw & \gate{U} & \qw & \gate{U} & \qw & \qw & \qw & \qw & \rstick{$\cdots$} \\
        \lstick{$\ket{0}$} & \qw & \gate{U} & \qw & \gate{U} & \qw & \qw & \qw & \qw & \rstick{$\cdots$}
    \end{quantikz}
    \begin{quantikz}[row sep=0.1cm, column sep=0.1cm]
        \lstick{$\ket{0}$} & \qw & \gate{U} & \ctrl{1} & \gate{U} & \qw & \targ{} & \qw & \qw & \rstick{$\cdots$} \\
        \lstick{$\ket{0}$} & \qw & \gate{U} & \targ{} & \gate{U} & \ctrl{2} & \qw & \qw & \qw & \rstick{$\cdots$} \\[.2cm]
        \lstick{$ \vdots $} & \wave &&&&&& \\ [.25cm]
        \lstick{$\ket{0}$} & \qw & \gate{U} & \ctrl{1} & \gate{U} & \targ{} & \qw & \qw & \qw & \rstick{$\cdots$} \\
        \lstick{$\ket{0}$} & \qw & \gate{U} & \targ{} & \gate{U} & \qw & \ctrl{-4} & \qw & \qw & \rstick{$\cdots$}
    \end{quantikz}
    \caption{Quantum circuits for the first experiments. The re-uploading model is depicted on top, and compared to the PQCs described in the bottom line. The $U$ gates here described correspond to arbitrary parameterized single-qubit operations. The circuit here described corresponds to one layer, and the depth is determined by the number of repetitions.}
    \label{fig.circuits}
\end{figure}
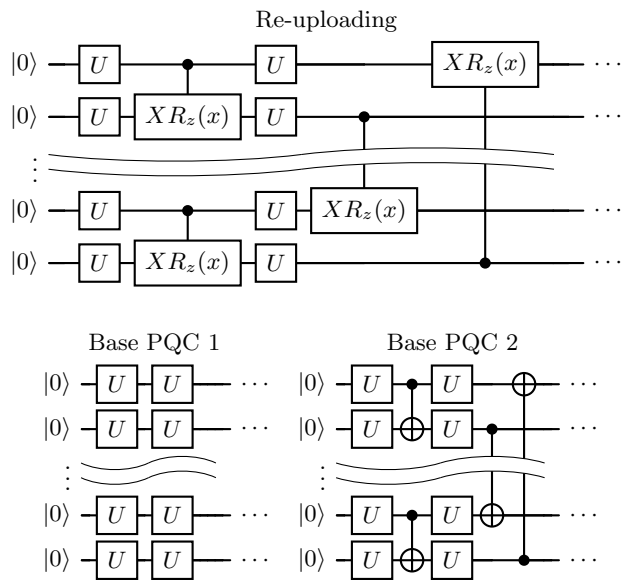

Consider again the previous example, this time with a different base PQC which includes entangling gates, see~\Cref{fig.circuits}, base PQC 2. The gates $U$ are considered distributed according to the Haar measure for single-qubit operations. In this new scenario, the PQC cost function $h_\btheta(0)$ suffers from BPs for sufficient depth~\cite{holmes2022connecting, cerezo2021cost}. Notice that it is possible to decompose a data-dependent entangling gate as a fixed entangling gate and tunable single-qubit~\cite{barenco1995elementary}, allowing for introducing data through single-qubit operations. Intuitively, the data can be re-absorbed by the parameters to generate a new circuit with the same ansatz as the base PQC. As a direct consequence, the gradients of the PQC and those of the QRU are of the same magnitude, $\Ex{\Var[\Btheta]{\partial_j h_\btheta(x)}}\approx \Var[\Btheta]{\partial_j h_\btheta(x_0)}$, for any $x_0$. This intuition motivates the newly coined concept of {\sl absorption witnesses} in~\Cref{def.absorption} as the capability of the circuit to absorb the data into the parameters.

\subsection{Absorption Witnesses}
Before further expanding on absorption witnesses, it is convenient to introduce some auxiliary quantities in the context of QRU models. We take derivatives with respect to the $j$-th parameter. All operations preceding the $j$-th parameter (not included) are considered the right part of the circuit, operationally attached to the input state $\rho_0$. Operations that include and follow the $j$-th parameter are on the left side of the circuit, attached to the observable. This description is given by
\begin{align}
    \rho_j(\btheta_{R, j}, x) & = U_{R, j}(\btheta_{R, j}, x) \ \rho_0 \ U^\dagger_{R, j}(\btheta_{R, j}, x) \\
    H_j(\btheta_{L, j}, x) & = U^\dagger_{L, j}(\btheta_{L, j}, x) \ H \ U_{L, j}(\btheta_{L, j}, x).
\end{align}
The left/right parameters $\Btheta_{R/L, j}$ are assumed to be independent. For each of the right and left parts of the circuit, we can define the difference with respect to the reference data value $x=0$ (corresponding to the PQC) as
\begin{align}\label{eq.absorption_diff1}
    B_{R, j}^{(t)}(\btheta_{R, j}, x; \rho_0) & = \rho_j^{\otimes t}(\btheta_{R, j}, x) - \rho_j^{\otimes t}(\btheta_{R, j}, 0) \\ 
    B_{L, j}^{(t)}(\btheta_{L, j}, x; H) & =    H_j^{\otimes t}(\btheta_{L, j}, x) -   H_j^{\otimes t}(\btheta_{L, j}, 0) \label{eq.absorption_diff2}
\end{align}
We define the absorption witness as follows.

\begin{definition}[Absorption witness]\label{def.absorption}
    Let $U(\btheta, x)$ be a re-uploading model as defined in~\Cref{eq.ru_model}. Let $U_{R/L, j}(\btheta, x)$ be the right and left parts of the circuit with respect to the $j$-th gate. 
    The right/left absorption witnesses are 
    \begin{align}
    \B^{(2)}_{R, j}(\rho_0) = & \mathbb E_X\left( \  \norm{\Er{B^{(2)}_{R, j}(\btheta_{R, j}, x;\rho_0)}}_1 \right), \\
    \B^{(2)}_{L, j}(H) = &  \mathbb E_X\left( \ \norm{ \El{B^{(2)}_{L, j}(\btheta_{L, j}, x;H)}}_1 \right).
    \end{align}
\end{definition}

The absorption witness defined above captures the effect of including data when averaging over the parameter space $\Btheta$. If $\B^{(2)}_{R, j}(\rho_0) = 0$, the input $x$ yields an effect on $\rho_j(\btheta_{R, j}, x)$ equivalent to some change $\btheta_{R, j}\rightarrow \btheta^*_{R, j}$. This effect is compensated when averaging over $\Btheta$. The logic is analogous to the left part of the circuit. As an illustrative example, assume a single-layer re-uploading model composed by applying any data-encoding layer after a PQC forming a $t$-design. By definition, $t$-designs approximate up to the $t$-th statistical moment of Haar measure and are thus insensitive (on average) to adding extra operations, in particular any operation given by data-encoding. However, this closeness to $t$-designs is no longer possible for ansatzes with (several) data-encoding gates interspersed between parameterized layers. 

The absorption witnesses from~\Cref{def.absorption} bound the differences between variances for PQCs and QRU models as follows.
\begin{theorem}\label{th.var_bounds}
    Let $U(\btheta, x)$ be a re-uploading model as defined in~\Cref{eq.ru_model}. Then 
    \begin{multline}\label{eq.var_bounds}
        \vert \Ex{\Var[\Btheta]{ \partial_j h_\btheta(x)}} - \Var[\Btheta]{\partial_j h_{\btheta}(0)} \vert  \\ \leq 4 \norm{V_j}_\infty^2 \left( \norm{H}_\infty^2\ \B^{(2)}_{R, j}(\rho_0) +\ \norm{\rho_0}_\infty^2\ \B^{(2)}_{L, j}(H) \right)
    \end{multline}
    where $\rho_0$ is the initial state, $H$ is the observable to measure, and $\B^{(2)}_{L, j}(H), \B^{(2)}_{R, j}(\rho_0)$ are the absorption witnesses from~\Cref{def.absorption}.
\end{theorem}
The proof can be found in~\Cref{app.var_bounds}.

Computing the absorption witness is not easy, nor computationally efficient. Alternatively, it can be estimated by comparing variances of the magnitudes of gradients with and without data, as will be done in the numerical calculations of~\Cref{sec.num_gradients}. Nevertheless, it provides a useful interpretation of the relationship between vanishing gradients for data-dependent QRU models, as compared to their base PQCs, where the BP phenomenon has been already studied~\cite{mcclean2018barren, larocca2022diagnosing, larocca2021theory}. The absorption witnesses quantify the expressivity difference between the PQC where the data uploading gates of the QRU are removed and that of the PQC where the data uploading gates are replaced by parameterized gates. As an illustrative example, consider an arbitrary Hamiltonian $H$, and a quantum re-uploading model composed of a gate $e^{iH\theta_0}$ immediately followed by a data uploading gate $e^{iHx_0}$. These gates can be combined as $e^{iH\theta_{1}}, \theta_1 = \theta_0 + x$. Since the relevant quantities are variances of, any shift of this kind does not affect the average behavior, yielding an absorption witness of exactly 0. On the other hand, a QRU model composed by two arbitrary Hamiltonians $e^{iH_1 \theta}$, $e^{i H_2 x}$ does not admit a shift in $\theta$ to absorb $x$, yielding an absorption witnwess depending on $H_{1, 2}$. This result is formulated in the same fashion as the ones in~\cite{holmes2022connecting}, extending their applicability to quantum machine learning. 

Finally, note that it is in principle possible to construct pathological datasets such that the base PQC suffers from BP while the QRU model does not. In these cases, the data uploading gates need a careful design to cancel out the structures responsible for vanishing gradients. It is thus reasonable to assume that real-world datasets would not result in such behavior.

\subsection{Gradients in layered QRU models}\label{sec.layered}

We consider in this section layered QRU models, in contrast to the results we presented earlier that apply to all QRU structures. In many practical scenarios there are several parameterized gates between each pair of encoding gates~\cite{larocca2022diagnosing, larocca2021theory}. An encoding gate and all preceding parameterized gates is referred to as a layer as
\begin{equation}\label{eq.layer_ru}
    U(\btheta, x) = \prod_{l = 1}^L  V_l(x) u_l(\btheta_l).
\end{equation}
In this representation, the parameterized gates $u(\btheta_l)$ are no longer defined by a single generator, and $\btheta_l$ is no longer one-dimensional. 
We can in this case study the absorption capability of each individual layer by defining the corresponding absorption witnesses as follows.
\begin{definition}[Layerwise absorption witness]\label{def.absorption_layer}
Let $u(\btheta_l)$ be the $l$-th layer of a re-uploading model from~\Cref{eq.ru_model}, and let $V(x)$ be the data-encoding operation applied immediately after $u(\btheta_l)$. The absorption witness for the $l$-th layer is
\begin{equation}
    \A_l^{(t)} = \E[X]{ \norm{\E[\Btheta_l]{V_l(x)^{\otimes 2} u(\btheta_{l})^{\otimes 2} - u_l(\btheta_{l})^{\otimes 2}}}_1 }.
\end{equation}
\end{definition}

We provide some examples where $\A_l^{(t)}=0$. First, assume a data-encoding layer sharing the generator with the corresponding parameterized gates. In this case, we can read data-encoding as a simple shift of parameters $\btheta^* \rightarrow \btheta - x$ (recall that $\btheta$ is now multi-dimensional), and averages do not change as long as $\btheta$ is sampled uniformly. Another example is the case where the ansatz is composed by $k$-local $2$-designs located in consecutively alternated qubits, as in~\cite{cerezo2021cost}, where any $k$-local data-encoding gates can be re-absorbed by definition. 

The use of layered ansatzes and layerwise absorption witnesses allows for further simplifications of~\Cref{th.var_bounds} by bounding the complete absorption witnesses.
\begin{lemma}\label{le.layers}
    Consider a layered re-uploading model as in~\Cref{eq.layer_ru}. Then
\begin{align}
    \B^{(2)}_{R, l+1}(\rho_0) & \leq \B^{(2)}_{R, l}(\rho_0) + \norm{\rho_0}_\infty^2 \A_{l+1}^{(2)} \\
    \B^{(2)}_{L, l}(H) & \leq \B^{(2)}_{L, l+1}(\rho_0) + \norm{H}_\infty^2 \A_{l}^{(2)}.
\end{align}
\end{lemma}
The proof can be found in~\Cref{app.layers}.

Consider now a layered circuit where $u_l(\cdot) = u_k(\cdot)$ for any pair $(l, k)$. The previous result can be further simplified since $\A_l^{(2)} = \A^{(2)}$ for all values of $l$. Therefore
\begin{corollary}
For a layered re-uploading model, the absorption witnesses of large parts of the circuits can be bounded by absorption witnesses of small pieces, by
\begin{align}
    \B^{(2)}_{R, l}(\rho_0) & \leq L \norm{\rho_0}_\infty^2 \A^{(2)}, \\
    \B^{(2)}_{L, l}(\rho_0) & \leq L \norm{H}_\infty^2 \A^{(2)}.
\end{align}
\end{corollary}
The proof follows by repeated application of~\Cref{le.layers}, together with the observation $\B^{(2)}_{R, l}(\rho_0) = 0$ if no data-encoding layer is considered in the absorption witness. We can therefore give a simplified bound for the results from~\Cref{eq.var_bounds} in the case of layered ansatz as
\begin{multline}\label{eq.bound_layered}
        \vert \Ex{\Var[\Btheta]{ \partial_j h_\btheta(x)}} - \Var[\Btheta]{\partial_j h_{\btheta}(0)} \vert  \\ \leq 8 L \norm{V_j}_\infty^2 \norm{H}_\infty^2\  \norm{\rho_0}_\infty^2 \A^{(2)}.
\end{multline}

The result from~\Cref{eq.bound_layered} is more loose than~\Cref{th.var_bounds}, but easier to compute, since it depends only on the layerwise absorption witness $\A^{(2)}$ corresponding to shallow circuits.

\subsection{Numerical results}\label{sec.num_gradients}

\begin{figure}[t!]
    \centering
    \includegraphics[width=\linewidth]{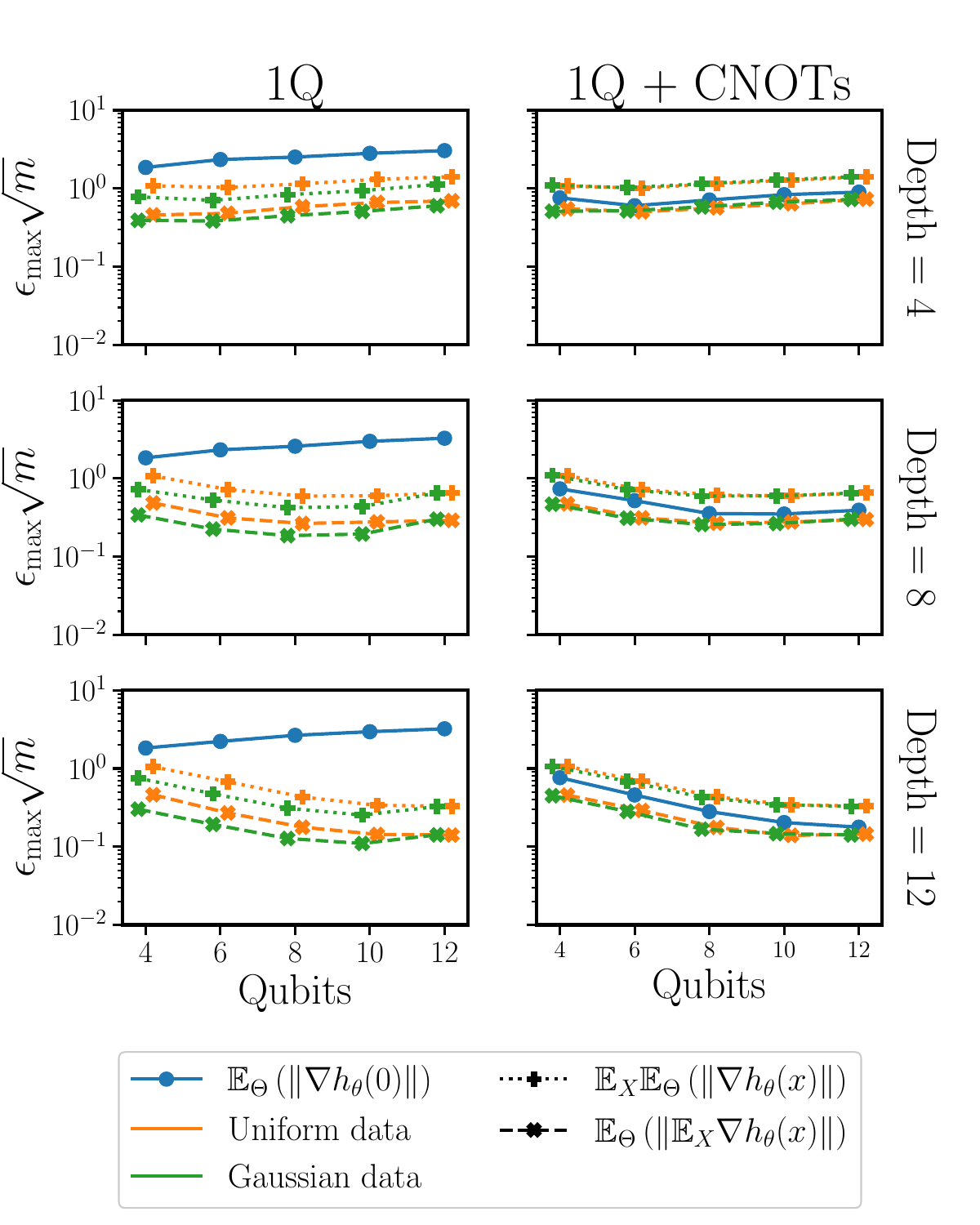}
    \caption{Results for
    $\epsilon_{\rm MAX} \sqrt{m} \approx \Ex{\E[\Btheta]{\norm{\nabla h_\btheta(x)}}}$ (see~\Cref{eq.ic}) for QRU models with alternating layered ansatzes. Data is introduced through controlled-rotation gates. Parameterized gates are single-qubit arbitrary operations. Each row has an increasing depth in the circuit. The right column includes CNOT gates for the base PQC. In the left column, gradients follow different trends for the cases with and without data, implying data can not be re-absorbed into a reparametrization, in the sense of~\Cref{th.var_bounds}. In the right column, similar trends indicate large absorption capabilities.
    }
    \label{fig.layered_ansatz}
\end{figure}
In this section, we present our numerical results, focusing on the average gradient magnitudes of hypothesis functions generated by QRU models in comparison to base PQCs. This analysis serves to validate the findings presented in ~\Cref{th.var_bounds} regarding gradient variances and can be considered as a proxy for evaluating the absorption witnesses defined in ~\Cref{def.absorption}. We explore various ansatzes and use different data distributions for the experiments. Our code for these experiments is available in~\cite{github}, and the data can be provided upon request.

For the numerical results we need to compute the magnitudes of the gradients on average. In order to reduce the computational complexity of this task, we will make use of the information content (IC) $I(\epsilon)$~\cite{perez-salinas2023analyzing}. The IC is a statistical measure of the variability of the optimization landscape.
In a nutshell, if $I(\epsilon)$ is close to $1$, then random displacements in $\bm\theta$ in the landscape change the value in $h_{\bm\theta}(x)$ in approximately $\epsilon$, conveniently re-normalized by the norm of the displacement itself. The value $\epsilon_{\rm MAX}$ at which $I(\epsilon)$ is maximized serves as a numerical proxy for the average norm of the gradient, that is
\begin{equation}\label{eq.ic}
    \E[\Btheta]{\norm{\nabla h_\btheta(x)}} \sim \epsilon_{\rm MAX} \sqrt m,
\end{equation}
where $\sqrt{m}$ is the number of parameters. While this approximation is not capable of computing the exact value of $\E[\Btheta]{\norm{\nabla h_{\btheta}(x)}}$, it is robust against statistical fluctuations and provides reliable scalings. We refer the interested reader to Ref.~\cite{perez-salinas2023analyzing} for an in-detail explaination of the validity and utility of IC to estimate gradients. 

\begin{figure}[t!]
    \centering
    \includegraphics[width=1\linewidth]{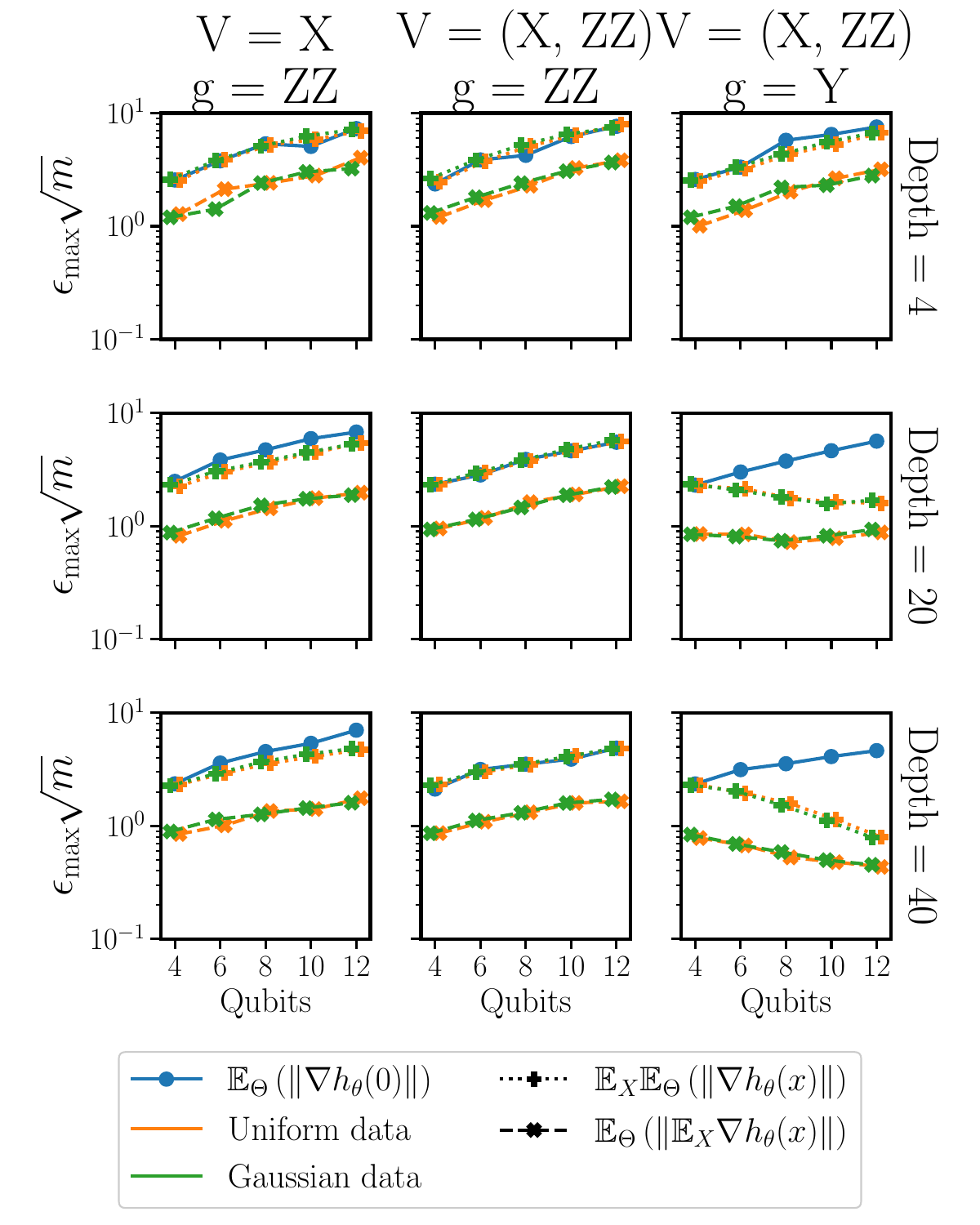}
     \caption{
     Results for
    $\epsilon_{\rm MAX} \sqrt{m} \approx \Ex{\E[\Btheta]{\norm{\nabla h_\btheta(x)}}}$ (see~\Cref{eq.ic}) for QRU models with translation-invariant layered ansatzes. Data is introduced through the generators $g$, and data through the generators $V$, indicated at the top for every column. Each row has an increasing depth in the circuit. In the left column, gradients follow approximately the same trends with and without data, implying high absorption capabilities in the sense of~\Cref{th.var_bounds}. For the middle column, the absorption is total since it can be done by a simple shift of parameters. The right column reveals different trends for the cases with and without data, implying low absorption. }
    \label{fig.transl_invariant_ansatz}
\end{figure}

As a first example, we compare a re-uploading ansatz, consisting of single-qubit rotations and data-encoding entangling gates, with two different PQCs (see~\Cref{fig.circuits}). In both cases, we construct the hypothesis function measuring sums of single-qubit $X$ Pauli measurements. The addition of entangling gates in PQC2 with respect to PQC1 is essential to exploring entangled states, and it plays a role in addressing the issue of vanishing gradients~\cite{cerezo2021cost}. The data-encoding layer is a controlled operation $C-(X R_z(x))$, which can be absorbed into single-qubit and controlled rotations~\cite{barenco1995elementary}. 

The results are shown in~\Cref{fig.layered_ansatz}. The columns correspond to the respective models (1-2), and the rows correspond to different depths of the ansatz. In the left column, corresponding to a PQC with only single-qubit gates, we observe a qualitative difference in the average gradients of the PQC and re-uploading circuits. The PQC is relatively insensitive to the number of qubits, as a consequence of the redundancy of multiple consecutive single-qubit gates. Adding data modifies this behavior. In the case of the 
entangling PQC (right column), the results without data align with previous works~\cite{cerezo2021cost, perez-salinas2023analyzing}. The addition of data drawn from different distributions (Gaussian and uniform) introduces negligible differences in the results.

We turn our attention to translation-invariant ansatzes. These circuits are not capable of freely exploring the Hilbert space, but only its invariant subspace. This restriction reduces the freedom in these circuits, leading to an increase in the average gradients of the cost function for PQCs~\cite{larocca2021theory}. We choose three layered models, based on the generators $X = \sum_q X_q$, $Y = \sum_q Y_q$, $ZZ = \sum_q Z_q Z_{q + 1}$, where $q$ cyclically iterates over all qubits. In the first model, the generator associated with parameters is $V_i = X$, and data-encoding is conducted through $g = ZZ$. The second model is given by $\{ V \} = \{X, ZZ\}$, $g = ZZ$. The third model is defined by $\{ V \} = \{X, ZZ\}$, $g = Y$.  In all cases, the observable considered is $X$. Among these models, only the second one can automatically absorb data into parameters through shifts. Gaussian-distributed data is used in all cases.

Results are detailed in~\Cref{fig.transl_invariant_ansatz}. The columns correspond to the respective models, and the rows correspond to different circuit depths. For each model, the average norm of the gradient scales differently with the number of qubits, with and without data. Models 1 and 2 present similar behavior when including data. In particular, for model 2 results show no difference between the re-uploading model and PQC since the data can be perfectly re-absorbed through a simple shift. A significant difference is noticeable in the third model. In this case, the absence of BPs in all instances makes the QRU models trainable by construction. 

\section{Expressivity in QRU models}\label{sec.expressivity}

The hypothesis class of QRU can always be expressed as a generalized trigonometric polynomial~\cite{caro2021encodingdependent}, see \Cref{eq.hyp_function_fourier}. In QRU models, the set of frequencies $\Omega$ is generated through the sequential Minkowski sum of the spectrum of the data encoding generators $\{\lambda_j\}_j$. In the general case, $\{\lambda_j\}_j$ consists of incommensurable real numbers, i. e. with non-rational ratios, and each new encoding step makes $\Omega$ combinatorially denser. In this section we first consider harmonic generators, i. e. with integer eigenvalues, and extend the results later to generic generators. As a main observation of this work, the behavior is similar in both cases.

\subsection{Harmonic representation of quantum states}\label{subsec.harmonicrep}

In this section, we introduce a representation of QRU models based on the Fourier decomposition of the hypothesis function. Such representation is useful for subsequent analytical results. Starting from~\Cref{eq.ru_model} and assuming the generators $g_i$ possess an integer spectrum, we can express the state before measurement as
\begin{equation}\label{eq.freq_repr}
    U(\btheta, x)\ket{0} =\sum_{k=-K}^{K} \sum_{j=1}^{2^n} c_{j,k}(\btheta) e^{i\mu kx}  \ket{j}.
\end{equation}
The coefficients $c_{j,k}$ form a matrix $\bm C \in \mathbb C^{2^n \times (2K + 1)}$ that defines uniquely (up to a global phase) the output state of the re-uploading circuit before measurement. The matrix $\bm C$ depends only on the parameters $\btheta$ and the generators of the ansatz, but not on the data $x$. The value $K$ corresponds to the largest attainable frequency, namely the sum of the largest eigenvalue for each generator used in the circuit. The recipe to construct $\bm C$ from the description of the circuit is detailed in \Cref{app.harmonic_qru}. Notice this approach is equivalent to adding an extra dimension (frequency)to the standard brute-force state vector simulation, which is not efficient from a computational point of view. This harmonic representation of QRU models simulator is available on~\cite{github}.

\subsection{Vanishing high frequencies in QRU models}\label{subsec.vanishinghf}
We use the above representations and the intuition that adding a data encoding layer corresponds to a convolution operation with the data encoding generator spectrum as defined below in \ref{def.spectrum}. For proof purposes, we assume that Haar random matrices are interleaved in between reuploading layers, as is common in most papers exploring barren plateaus. We examine the statistical properties of the amplitude of the coefficients as a function of the frequency. We begin by defining the spectrum kernel of a harmonic Hermitian matrix. 
 
\begin{definition}[Harmonic spectrum kernel]\label{def.spectrum}
    Let $H$ be a $N \times N$ Hermitian matrix with integer eigenvalues $\{\lambda\}$ with multiplicities $m(\lambda)$. The spectrum kernel of $H$ is the vector (indexed by $k$)
    \begin{equation}
        \K_H(k) = \left\{ 
        \begin{matrix}
            m(k \mu) / N & \textrm{if $k\mu  \in \{\lambda\}$} \\ 
            0 & {\rm Otherwise}
        \end{matrix}\right. ,
    \end{equation}
    where $\mu$ is the largest value in $\mathbb R$ compatible with this description.
\end{definition}
This function simply maps the eigenvalues of a Hermitian matrix into the normalized dimensionality of the corresponding eigenspace. For readability, we will refer to the {\sl spectrum multiplicity function} simply as the {\sl spectrum} for the remainder of the paper.

In the case of layered QRU models, the spectra of their data-encoding generators and the number of layers $L$ directly determine the set of attainable frequencies. The maximum attainable frequency is bounded by $L \norm{g}_2 $. We provide now some insight into how the coefficients are expected to behave. 
\begin{lemma}[Harmonic convolution]\label{le.harmonic_spectrum}
Let $\ket{\psi_\btheta(x)} = U(\btheta, x) \ket{\psi_0}$ be the output state of a re-uploading model, with data-encoded through the generator set $\{g_j\}$, each with spectrum $\K_{g_j}$, encoded as in~\Cref{eq.freq_repr}. Assuming that each parameterized step is drawn from the Haar measure of unitaries, then $\sum_{j} \abs{c_{j,k}}^2$ is a random variable satisfying
\begin{multline}
    \sum_{j} \abs{c_{j,k}}^2 \sim \\ \Dir\left(\left(\K_{g_1} \ast \ldots \ast \K_{g_j} \ast \ldots \ast \K_{g_L}\right)(k)\right),
\end{multline}
where $\Dir(\alpha_1, \alpha_2, \ldots )$ is the Dirichlet distribution~\cite{olkin1964multivariate} and $\ast$ denotes the convolution. 
\end{lemma}
The proof can be found in~\Cref{app.conv_spectrum}. The Dirichlet distribution is a family of probability distributions for multidimensional variables $\bm x \in [0, 1]^N$, subject to $\Vert \bm x \Vert_1 = 1$. The Dirichlet distribution over $N$ variables is fully described by $N$ parameters $\alpha_i \in \mathbb R_{> 0}$. Dirichlet is the multidimensional extension of the beta distribution. A detailed definition of the Dirichlet distribution and auxiliary results are given in~\Cref{app.dirichlet}. 
For completeness, we define convolution as
\begin{equation}
    (f * g)(k) = \sum_{l = -\infty}^\infty f[k] g[l - k].
\end{equation}
In other words, this lemma gives statistical properties of the frequency content, expressed as the norm of the $2^n$ quantum vector corresponding to each frequency (seee \Cref{eq.freq_repr}) for QRU models composed of a sequence of data uploading gates interleaved with gates drawn from the Haar distribution. It states that the vector of frequency content follows a multidimensional distribution whose mean is the result of the successive convolution of the multiplicity kernels of the data encoding gates. It follows a Dirichlet distribution because all values are positive and sum up to one as per the normalization of a quantum state.

It is worth discussing the role of the Haar distribution in this result. First, choosing random unitaries allows us to scramble the inner quantum state in the QRU model at each step, thus transforming the QRU circuit into a random walk in the space of frequencies, where the parameters in Dirichlet only account for the number of paths leading to the same outputs. Second, random choices of unitaries is in alignment with other works exploring trainability and expressivity in VQAs~\cite{mcclean2018barren, cerezo2021cost, holmes2022connecting}, which rely on sampling unitaries from a $t$-design. The difference between the Haar distribution and a $t$-design is rather technical, since $t$-designs are sets of unitaries with the same statistical moments as the Haar measure, up to degree $t$~\cite{sim2019expressibility}. With respect to~\Cref{le.harmonic_spectrum}, lowering the requirements in the parameterized steps from Haar distribution to $t$-design would imply to substitute the Dirichlet distribution with another probability distribution with the same $t$-statistical moments. Technical descriptions of this transformations are left as open questions for future research. Note that our results lose their validity if the parameterized steps are drawn with respect to other distributions of unitaries. 

The previous result immediately implies the following. 
\begin{theorem}[Single-generator convolution]\label{th.conv_spectrum}
    Let $\ket{\psi_\btheta(x)} = U(\btheta, x) \ket{\psi_0}$ be the output state of a re-uploading model, with data-encoded through the generator $g$, with spectrum $\K_{g}$, encoded as in~\Cref{eq.freq_repr}. 
     Assuming that each parameterized step is drawn from the Haar measure of unitaries, then 
\begin{equation}
    \sum_{j} \abs{c_{j,k}}^2 = \Dir\left(\left(\K_{g}^{*L}\right)(k)\right),
\end{equation}
where $(\cdot)^{*L}$ denotes the $L$-fold convolution.
\end{theorem}
The proof is immediate from extending~\Cref{le.harmonic_spectrum}.

We provide two explicit examples to distinguish the cases captured by \Cref{le.harmonic_spectrum} and \Cref{th.conv_spectrum}. Consider the single-qubit generator $g = (Z_0 + I)/2$, with spectrum $\K_g = (0, 1)$. To illustrate \Cref{le.harmonic_spectrum}, we choose the list of generators as $\{2^l g\}_{l = 0}^L$, yielding a convolution
\begin{multline}
    \left(\K_{g_1} \ast \ldots \ast \K_{g_j} \ast \ldots \ast \K_{g_L}\right)(k) = 1, \\ \forall k \in \{0, \ldots, 2^L - 1\}. 
\end{multline}
On the other hand, illustrating \Cref{th.conv_spectrum} we consider a repeated application of $g$, yielding
\begin{equation}
    \left(\K_g^{\ast L}\right)(k) = \binom{k}{L}, \forall k \in \{0, \ldots, L\}. 
\end{equation}
The behavior in the two cases of the random variable $\sum_{j} \abs{c_{j,k}}^2$ is significantly different. In the first case, the output is a flat distribution of exponential size. On the contrary, the second case is a distribution of linear size with high concentration in its mean values. 

Notice that, provided that the data generator is known, it is possible to classically store $\K_g^{* L}$ within memory of size $\bigO{L\norm{g}_2 / \mu}$, with computational cost $\bigO{(L\norm{g}_2 / \mu)^3}$. This allows us to classically characterize the frequency profile prior to executing the QRU model in quantum hardware for harmonic generators with only polynomially many eigenvalues.

The previous theorem can be readily interpreted in the limit of large $L$ by virtue of the central limit theorem~\cite{billingsley1995probability}. The repeated convolution of any random variable with a variance of $\sigma$ and a probability distribution in the spaces $L^1$ and $L^2$, tends to a normal distribution in a weak sense. We can thus obtain the following result.
\begin{corollary}[Vanishing high frequencies]\label{cor.gaussian_approx}
    In the conditions of~\Cref{th.conv_spectrum} and for large number of re-uploading $L$, 
    \begin{equation}\label{eq.gaussian_approx}
    \lim_{L \rightarrow \infty}\sum_{j} \abs{c_{j,k}}^2 \sim \Dir\left(\N\left(0, \sigma_g^2 L\right)(k)\right),
\end{equation}
where $\sigma_g$ is the standard deviation of the spectrum $\K_g$, and $\N(\mu, \sigma^2)$ is the normal distribution.
\end{corollary}
This observation implies that tails of the distribution vanish exponentially for large frequencies and there is a concentration in the low-frequency terms as the magnitudes of high-frequency terms vanish. In asymptotic scaling the available spectrum reduces from $\norm{g}_\lambda^2 L$ to $\sigma_g \sqrt{L}$. For interpretability, recall the example $g = (Z_0 + I)/2$, yielding a binomial distribution in the convolution of the subsequent spectra. The binomial distribution rapidly tends to a gaussian distribution. 

The results from~\Cref{th.conv_spectrum} and~\Cref{cor.gaussian_approx} can be extended to non-harmonic generators. For readability, we postpone this result until~\Cref{sec.nonharmonic}. 

The previous discussion considers the effects of the spectrum on the internal state of the re-uploading model in its harmonic representation. We are however not interested in the state itself, but rather in $h_\btheta(x)$ measured as an expectation value of this internal state. The Fourier components $h_\btheta(x)$ satisfy the following corollary.

\begin{corollary}\label{cor.observable}
    Let $\ket{\psi_\btheta(x)}$ be the output state of a re-uploading model, with a single data-encoding generator $g$ with spectrum $\K_g$. Let $h_\btheta(x)$ be the hypothesis function induced by the observable $H$ in the re-uploading model, as in~\Cref{eq.hypothesis_function}, and let $a_k(\btheta)$ be their corresponding Fourier coefficients as in~\Cref{eq.hyp_function_fourier}. In the conditions of~\Cref{th.conv_spectrum}, and for symmetric spectra $\K_g(k) = \K_g(-k)$,
    \begin{align}
     \norm{H}_\lambda^{-2} \left\vert a_k(\btheta) \right\vert^2 & \leq p_k \\
     p_k & \sim \Dir(\K_g^{*2L}(k)),\label{eq.dirichlet}
\end{align}
where $p_k$ is a multidimensional probability distribution sampled from the Dirichlet distribution defined by the $L$-fold convoluted spectrum $\Dir(\K_g^{*2L}(k))$~\cite{olkin1964multivariate}.
\end{corollary}
Additionally, this result extends to the Gaussian distribution in the limit of large $L$ as for~\Cref{cor.gaussian_approx}. 

The results from this section show that the frequency terms of $h_\theta(x)$ tend to follow a Gaussian profile of width $\sim L$, in the assumption that the generator of data-encoding gates is repeated in the QRU model. However, the frequency support of these functions scales linearly in $L$. As an immediate consequence, only frequencies $\omega \in \mathcal{O}(\sqrt{L})$ have practical support on average, while larger frequencies have exponentially vanishing weight in the hypothesis function. Note, that the Gaussian profile described by~\Cref{cor.gaussian_approx} does not imply a dense frequency space, which is still restricted to integer frequencies. This result holds even in the case where the generator provides exponentially-in-qubits many frequencies. It is then possible to have exponentially large frequency sets even with a small number of re-uploading steps, and the Gaussian approximation still holds with $ \sigma \sim \sqrt L e^n$.

\subsection{Lipschtitz expressivity}\label{subsec.lipschitz}

In this section, we delve into a more practical understanding of the expressivity of hypothesis functions in terms of the magnitude of their derivatives. The ability to capture fine-grained data patterns depends on the function's ability to access high rates of change, i.e., the magnitude of its derivative. This concept can be quantified through the maximum value of the derivative, known as the optimal Lipschitz constant. For a function $f$, the optimal Lipschitz constant is defined as
\begin{equation}\label{eq.lipschitz}
    \L(f) = \max_x \left\vert \partial_x f(x)\right\vert.
\end{equation}
The Lipschitz constant is closely related to Fourier analysis, as high derivatives can only be achieved if the Fourier spectrum includes high frequencies with significant coefficients. Specifically,
\begin{equation}
    \L(f) \leq \sum_{k = -K}^{K} \vert k \vert \mu \vert a_k e^{ik\mu x} \vert,
\end{equation}
where $a_k$ represents the Fourier coefficients of the hypothesis function.

We introduce an upper bound to the Lipschitz constant inspired by~\Cref{eq.lipschitz}, adapted for QRU models and properly normalized with respect to the measured observable as
\begin{equation}\label{eq.lipschitz_def}
    \Lambda(h_\btheta) = \sum_{k = -K}^K \mu \vert k \vert \vert a_k \vert.
\end{equation}
It is straightforward to see that $\Lambda(h_\btheta) \geq \L(h_\btheta)$, and therefore we are going to use this quantity as a proxy for it. For readability, this optimal Lipschitz constant upper bound will be referred to LB in the subsequent sections of this paper and be noted $\Lambda(h_\btheta)$ unless otherwise specified.

Using results from previous sections we study $\Lambda(h_\btheta)$, starting with a result giving tight bounds on the asymptotic average of the LB over the parameters $\Btheta$. The results stated in the next and subsequent propositions stem from the conditions discussed in~\Cref{subsec.vanishinghf}, namely tunable gates are drawn from the Haar distribution.  
\begin{theorem}[Average LB]\label{th.av_lipschitz}
Let $h_\btheta(x)$ be the hypothesis function of a re-uploading model for which~\Cref{th.conv_spectrum} applies. Let $\Lambda(h_\btheta)$ be the LB as defined in~\Cref{eq.lipschitz_def}. Then, 
    \begin{equation}
       \norm{H}_\lambda \sqrt{2 L} \mu\sigma_g \leq \lim_{L \rightarrow \infty}\E[\Btheta]{\Lambda(h_\btheta)} \leq  \norm{H}_\lambda \frac{4}{\sqrt{\pi}} \sqrt{L}  \mu\sigma_g 
    \end{equation}
\end{theorem}
The proof can be found in~\Cref{app.av_lipschitz}. Notice the tightness of the bounds above since $2\sqrt{2} / \sqrt{\pi} \approx 1.6$.

The following subsection quantifies the likelihood of the LB different from the average. Notice that values smaller than the average are not relevant due to the definition of $\Lambda(h_\btheta)$. We can leverage the insights from previous results, particularly the role of Dirichlet distributions, to derive the following result:

\begin{theorem}[Deviation of LB]\label{th.dev_lipschitz}
Let $h_\btheta(x)$ be the hypothesis function of a re-uploading model for which~\Cref{th.conv_spectrum} applies, with data-encoding generator $g$. Let $\Lambda(h_\btheta)$ be its LB as defined in~\Cref{eq.lipschitz_def}. Then, 
        \begin{multline}
    \lim_{L \rightarrow \infty} \operatorname{Prob}\left( \Lambda(h_\btheta) - \norm{H}_\lambda \sqrt{2 L }\sigma_g\mu\geq t \right) \\ \in \mathcal{O}\left( \exp \left(-\frac{t^2}{{\rm poly}(L \mu)} \right) \right).
\end{multline}
\end{theorem}
The proof can be found in~\Cref{app.dev_lipschitz}. Notice this result automatically bounds the probability of the optimal Lipschitz constant itself of being bigger than $\sqrt{2 L \mu} \sigma_g$.

The previous theorem can be further refined to provide a tighter bound on the likelihood of large deviations from the LB. As mentioned in the detailed proof, when Theorem \ref{th.conv_spectrum} holds the weight of each frequency and tends to follow a Gaussian-like profile, with central frequencies having exponentially larger probabilities than the extremal ones. It is expected that the primary contributions to $\Lambda(h_\btheta)$ come from these central frequencies, which also have the smallest prefactors.
Taking this into account, we can update the results from Theorem \ref{th.dev_lipschitz} to provide a more precise bound,
\begin{multline}\label{eq.tight_bound_lipschitz}
    \lim_{L \rightarrow \infty} \operatorname{Prob}\left( \Lambda(h_\btheta) - \norm{H}_\lambda \sqrt{2 L }\mu\sigma_g\geq t \right) \\ \in \mathcal{O}\left( \exp \left(-\frac{t^2}{(\sigma_g \mu\sqrt{L})^3} \right) \right).
\end{multline}

The vanishing high frequencies from~\Cref{subsec.vanishinghf} have consequences on the properties of the attainable hypothesis functions. In particular, its maximal derivative with respect $x$, given by the Lipschitz constant, scales in average with $\sqrt L$, and the probability of finding larger Lipschitz constants vanishes super-exponentially fast. This imposes in practice constraints on the capability of the hypothesis functions to capture fine details in the data, effectively restricting target functions that can be approximated by QRU models.

\subsection{Extension to generic data generators}\label{sec.nonharmonic}
In previous subsections, we have proven the phenomenon of vanishing high frequencies and its consequences on the Lipshitz constant for harmonic data generators. In this subsection, we extend the results from~\Cref{subsec.vanishinghf} and~\ref{subsec.lipschitz} to any data generator. We start by defining the spectrum kernel for generic Hermitian matrices.
\begin{definition}[Hermitian spectrum kernel]\label{def.spectrum_nonharmonic}
    Let $H$ be a $N \times N$ Hermitian matrix with integer (positive or negative) eigenvalues $\{\lambda\}$ with multiplicities $m(\lambda)$. We define the vector $\vec \mu \in \mathbb R^D$, with $\mu_i / \mu_j \in \mathbb R \backslash \mathbb Q \ \forall(i,j)$, such that any eigenvalue can be written as $\lambda = \vec \mu \cdot \vec k$, with $\vec k \in \mathbb Z^D$. We refer to the number of anharmonic dimensions as $D \leq 2^n$, where $n$ is the number of qubits. We define the spectrum kernel of $H$ as $\K_H$ such that
    \begin{equation}
        \K_H(\vec k) = \left\{ 
        \begin{matrix}
            m(\lambda) / N & \textrm{if $\vec k \cdot \vec \mu  \in \{\lambda\}$} \\ 
            0 & {\rm Otherwise}
        \end{matrix}\right.
    \end{equation}
    Where each $\mu_j$ is the largest value in $\mathbb R$ compatible with this description.
\end{definition}

We note the covariance of this spectrum as the $D\times D$ matrix $\Sigma_g$. The average of this spectrum is $0$ since we consider traceless generators. The results from~\Cref{le.harmonic_spectrum} and~\Cref{th.conv_spectrum} hold in the non-harmonic case. In this scenario, the convolution must be done in a $D$-dimensional space, leading to $D$-dimensional frequency profiles. The convoluted spectrum, for the single-generator case, can be stored in a memory structure of size $\bigO{\left(L\norm{g}_2 / \min_j{\mu_j}\right)^D}$. Notice that for each eigenvalue $\lambda$ there exists only one compatible $\vec k$, due to the irrational ratios between elements in $\vec \mu$.

The central limit theorem still applies in the non-harmonic case as well, leading to the following result. 

\begin{corollary}[Vanishing high frequencies]\label{cor.gaussian_approx_nonharmonic}
    Given the conditions of~\Cref{th.conv_spectrum} for non-harmonic generators and for large number of re-uploadings $L$, 
    \begin{equation}\label{eq.gaussian_approx_nonharmonic}
    \lim_{L \rightarrow \infty}\sum_{j} \abs{c_{j,\vec k}}^2 \sim \Dir\left(\N\left(0, \Sigma_g L\right)\left(\vec k\right)\right).
\end{equation}
\end{corollary}

Following the reasoning from the harmonic case, we focus now on the Lipschitz constant of the hypothesis functions. The definition from~\Cref{eq.lipschitz_def} can be extended to
\begin{equation}\label{eq.lipschitz_def_nonharmonic}
    \Lambda(h_\btheta) = \sum_{\omega \in \Omega} \vert \omega \vert \vert a_\omega \vert,
\end{equation}
with $\omega = \vec \mu \cdot \vec k$. Since $\vec k$ has integer values and $\vec\mu$ has irrational ratios among its elements, there is at most one solution of $\vec k$ for each $\omega$. With this definition, we can formulate results analogous to~\Cref{th.av_lipschitz} and~\Cref{th.dev_lipschitz}.

\begin{corollary}[Lipschitz bounds for non-harmonic generators]\label{cor.lipschitz_nonharmonic}
Let $h_\btheta(x)$ be the hypothesis function of a re-uploading model for which~\Cref{cor.gaussian_approx_nonharmonic} applies. Let $\Lambda(h_\btheta)$ be the LB as defined in~\Cref{eq.lipschitz_def_nonharmonic}. Then, 
\begin{align}\label{eq.av_lipschizt_nonharmonic}
    \lim_{L \rightarrow \infty}\E[\Btheta]{\Lambda(h_\btheta)} & \leq  \frac{4}{\sqrt \pi}\norm{H}_\lambda \sqrt{\Tr(\Sigma)} \ \norm{\vec \mu}_2 \sqrt{L} \\
    \lim_{L \rightarrow \infty}\E[\Btheta]{\Lambda(h_\btheta)} & \geq \sqrt 2 \ \norm{H}_\lambda \sqrt{\min_\lambda(\Sigma)} \, \norm{\vec\mu}_2 \sqrt{L} .
\end{align}
\end{corollary}

The proof of~\Cref{cor.lipschitz_nonharmonic} can be found in~\Cref{app.nonharmonic}. In addition, following the same reasoning leading to~\Cref{th.dev_lipschitz}, we can infer exponential concentrations of $\Lambda(h_\btheta)$ around its average values, by
\begin{multline}
    \label{eq.tight_bound_lipschitz_nonharmonic}
    \lim_{L \rightarrow \infty} \operatorname{Prob}\left( \Lambda(h_\btheta) -  \norm{H}_\lambda \sqrt{2 L} \norm{\vec\mu}_2 \sqrt{\min_\lambda(\Sigma)}\geq t \right) \\ \in \mathcal{O}\left( \exp \left(-\frac{t^2}{\left(\sqrt{\max_\lambda(\Sigma_g)}\max_j(\mu_j)\sqrt{L}\right)^3} \right) \right).
\end{multline}

In light of the previous theorem, we can observe that the vanishing high frequencies phenomenon extends to non-harmonic generators, with minor changes with respect to the harmonic case, rooting from norm bounds in the multi-dimensional space. The tightness of these bounds depends on the regularity of the anharmonic spaces, which is reflected into the values of $\vec\mu$ and the eigenvalues of $\Sigma$. 

An immediate consequence of this section is that QRU models can have a dense frequency spectrum without significantly modifying the envelope of the frequency profile. The only elements of QRU models allowing to increase the set of available frequencies in practice are $L$ and the spectrum profile $\Sigma$, while $\vec\mu$, which can be related to $\norm{g}_\lambda$ have a more modest effect. It is possible to reach exponentially many different frequencies by using generators with exponentially large $\K_g$.

The number of different frequencies directly affects the surrogability of the studied QML models In the case the frequency space is polynomial in the number of qubits, it is possible to construct a classical model fitting the corresponding generalized trigonometric polynomial~\cite{schreiber2023classical}. On the other hand, exponentially large frequency spaces do not admit arbitrary efficient classical representations. The findings detailed in this work provide methods to circumvent surrogability. This can come from generators with exponentially large spectra, or designed in such a way that the frequency space scales exponentially with $L$, for instance with convolutions of highly non-harmonic spectra. 

\subsection{Numerical results}\label{sec.num_expressivity}
In this section, we show the results of a series of numerical experiments in which such conditions are relaxed and show that the theoretical results still apply. We use three models to test different situations. The first two models are constructed with permutation-invariant generators, which correspond to PQC that have been proven to be trainable~\cite{schatzki2022theoretical}. Those models express only the symmetric subspace in the available Hilbert space. In the first model (A), $g = X, V = ZZ$, and the second model (B) $g = X, V = \{Y, X, ZZ\}$. For the third model, $g = X$, and the parameterized pieces are sampled from the Haar measure, that is the set $V$ is free. We choose these models to have full control of the spectrum of the generator $\K_{g = X}$, which allows us to informatively compare to the theorems. All experiments were conducted with systems of $4$ qubits unless explicitly stated, without affecting the scaling of the obtained results. 
 
\begin{figure}[t!]
    \centering
    \includegraphics[width=1.075\linewidth]{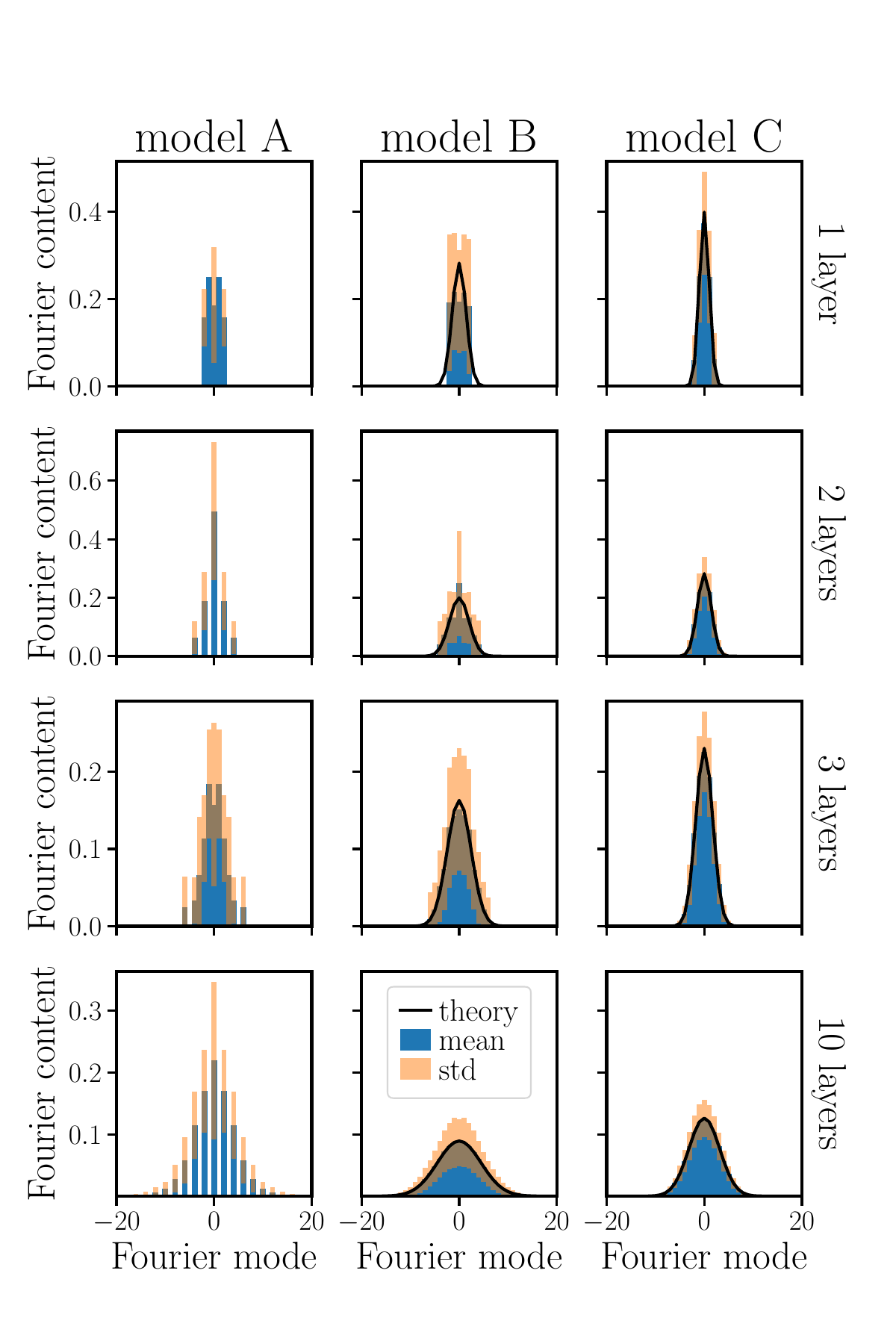}
    \caption{Evolution of frequency spectrum with the number of layers, for models (A, B, C) detailed in the first paragraph of the section. The Fourier content refers to (average and standard deviation of) $\vert c_k \vert^2$. Model (A) is not general enough to follow~\Cref{th.conv_spectrum}, but still, a spread in frequencies is observed. Model (B) is permutation-invariant and almost fully general in the symmetric space, and model (C) is general with no restrictions in the Hilbert space. As $L$ increases, the Fourier spectrum approximates a Gaussian profile with increasing variance according~\Cref{eq.gaussian_approx}. The values $\sigma_g$ for models (B) and (C) change due to the constraint in the available space.}
    \label{fig.kernels_circuit}
\end{figure}
\begin{figure}[t!]
    \includegraphics[width=\linewidth]{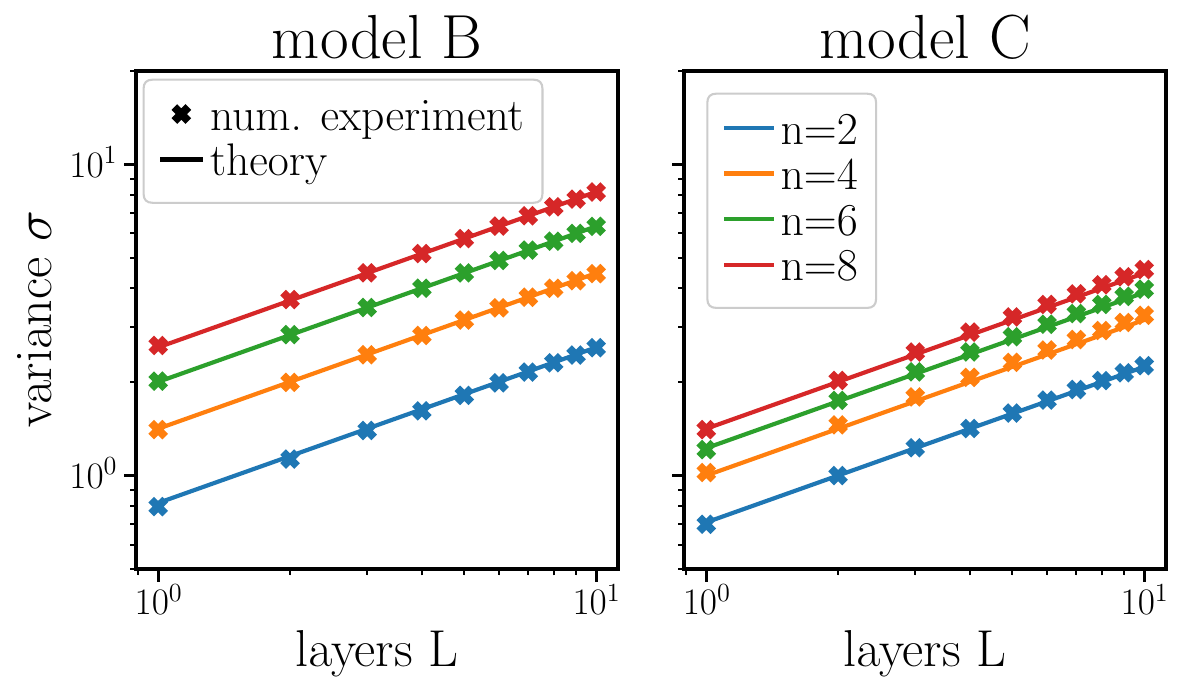}
    \caption{Variances for the tending-to-Gaussian profiles from~\Cref{fig.kernels_circuit}, for different numbers of qubits, in the y-axis, while the x-axis corresponds to the number of layers. The left figure corresponds to model (B), with $\sigma_g^2 = n(n+2) / 12$. The right figure corresponds to model (C), with $\sigma_g^2 = n / 4$. The scaling in $\sqrt{L}$ is in agreement with~\Cref{eq.gaussian_approx}.}
    \label{fig.kernels_variance}
    \includegraphics[width=\linewidth]{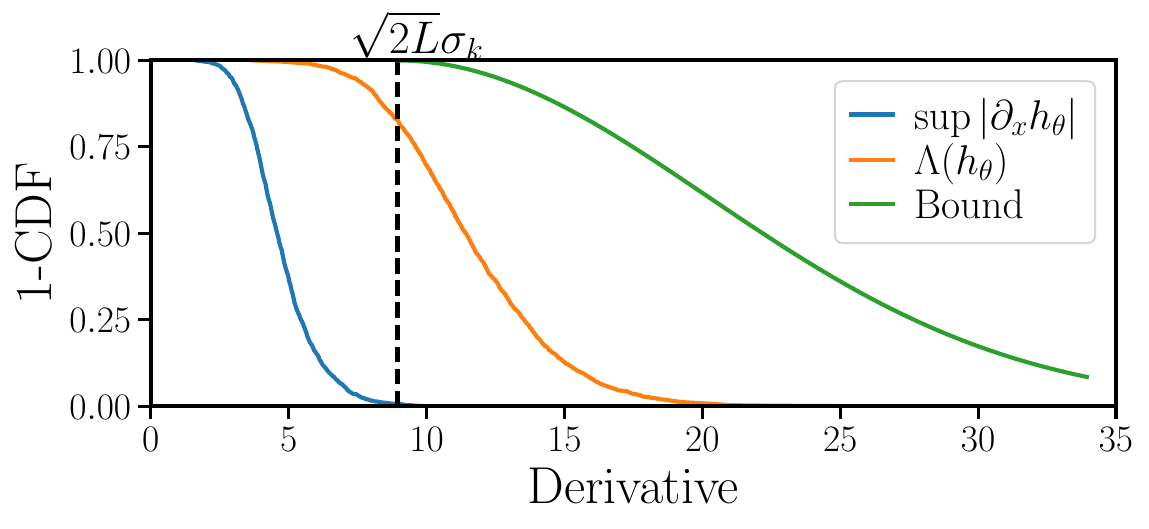}
    \caption{Inverse CDFs for the optimal Lipschitz constants and $\Lambda(h_\btheta)$, as compared to the bounds from~\Cref{th.dev_lipschitz}. The x-axis indicate the value for each CDF, respectively $\sup_x \vert \partial_x h_\btheta(x)\vert$, $\Lambda(h_\btheta)$ and $t$ for each line. }
    \label{fig.lipschitz}
\end{figure}

The first experiment tests~\Cref{th.conv_spectrum} and~\Cref{cor.gaussian_approx}, and the results can be seen in~\Cref{fig.kernels_circuit}. In model (A), the spectrum spreads towards large frequencies with the number of data re-uploadings. The parameterized gates are not general enough to support the theoretical results. For models (B) and (C), the Gaussian limit is matched even for a moderately small number of layers. This means that even though the theorem is proven for Haar random unitaries, the vanishing high-frequencies behavior still holds for model (B), even though the Haar condition is not guaranteed. Notice the difference in spreads for models (B) and (C). This is a consequence of the space explored by the ansatz. Model (B) is composed of a permutation invariant ansatz, and it is as general as possible only in the symmetric subspace, of dimension $n+1$. In this scenario, the spectrum of the corresponding $g$ is flat (see the results for 1 layer in figure \ref{fig.kernels_circuit}), and the spread depends on the number of qubits $n$ as $\sigma_g = \bigO{n}$. For the model (C), the spectrum of $g$ with no restriction follows a binomial distribution, centered in $k = 0$, with $\sigma_g \in \bigO{\sqrt n}$. A comparison between the theoretical and observed variances can be found in~\Cref{fig.kernels_variance}, showing high agreement with the theoretical results.

We numerically check~\Cref{th.dev_lipschitz} in~\Cref{fig.lipschitz}. We depict in this figure the observed cumulative distribution functions (CDF) of both the numerically found LB, and  $\Lambda(h_\btheta)$ defined in~\Cref{eq.lipschitz}. These CDFs are compared to the upper bound from~\Cref{th.dev_lipschitz}. Results show agreement with~\Cref{th.dev_lipschitz}, and even indicate the possibility of finding tighter bounds, at least in terms of prefactors.

\subsubsection{Training }
All the LB results describe an average behavior for $\Btheta$. In this subsection, we briefly explore the effect of previous results in the training. We task model (B) to learn functions whose Fourier coefficients follow a step function of increasing width. This approximately corresponds to a cardinal sinus of decreasing width. 

We display results of trained QRU models in~\Cref{fig.training_VHF}. In the top figure, we show the Fourier components of different functions to be fitted (in dashed lines), and the hypothesis functions after training (solid lines). The target function is learned by the model for $K \leq 20$ but the hypothesis function fails to capture high frequencies from $k > 25$. Notice that the obtained hypothesis functions for $K = \{30, 40\}$ seem to saturate the expressivity capabilities of the model. The bottom figure represents the functions in the data domain for $K = \{4, 40\}$.

\begin{figure}[t!]
    \includegraphics[width=1\linewidth]{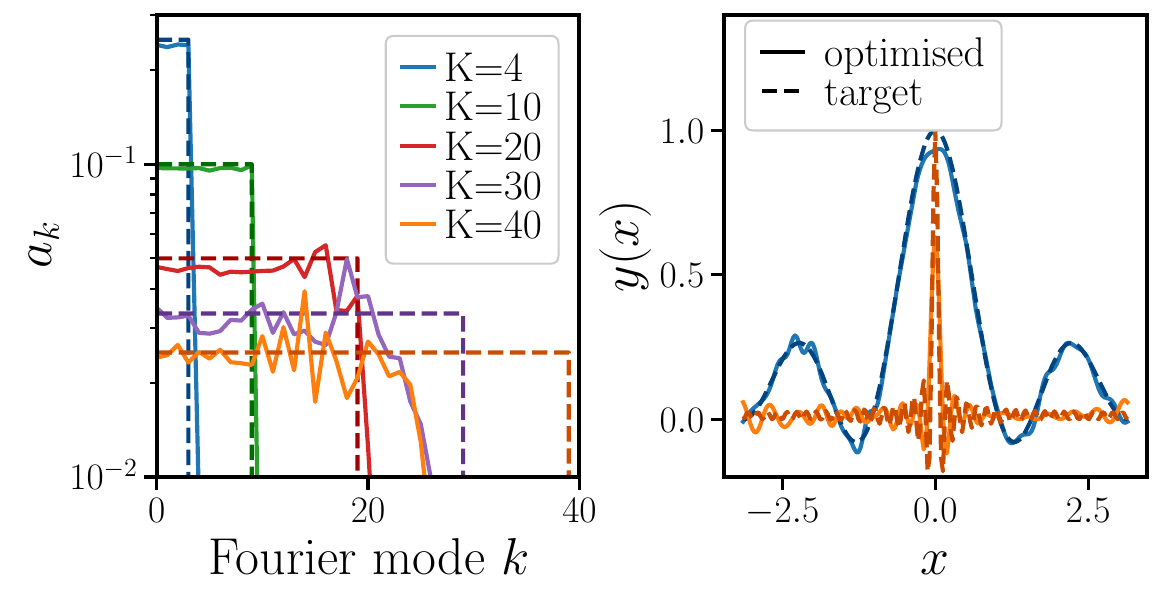}
    \caption{Time and frequency domains representations of the functions of circuits (B) trained to match increasingly sharp cardinal sinus that yields increasing high-frequency content. From Fourier mode $k=20$, the model is not able to match the amplitude, exhibiting a consequence of vanishing high frequencies.
    \label{fig.training_VHF}}
\end{figure}

\section{Discussion}\label{sec.discussion}

We turn our attention first to gradients. From our results, we can infer that vanishing gradients are avoided in QRU models if the base PQC is BP-free, and data can be absorbed into the parameterized gates. We can refer to existing literature on avoiding BPs for PQCs by restricting the dimensionality of the search space, by means of the dynamical Lie algebra~\cite{larocca2021theory, schatzki2022theoretical}. In a nutshell, the Lie algebra depends on the generators of the quantum model. Absorption witnesses can only be maintained close to $0$ if the base PQC and the derived QRU model share a common Lie algebra. This observation allows one to choose data-encoding generators avoiding the emergence of BPs. 

The average of $\Lambda(h_\btheta)$ is a consequence of the vanishing high frequencies behavior that grows as $\sim  \sigma_g\sqrt{L}$, as imposed by the central limit theorem. Deviating from this average is exponentially unlikely, as proven in theorem \ref{th.dev_lipschitz}. As discussed later, it is in principle possible to amplify high-frequency components, at the expense of losing all degrees of freedom in the process. Therefore, for practical purposes, we need to adjust the number of re-uploading layers according to the scaling $\sim \sqrt L$, and not $\sim L$, as suggested by other theoretical works on expressivity via generalization bounds~\cite{caro2021encodingdependent}. 

In this work we derived the the scaling of the Fourier spectrum of hypothesis functions with the number of layers, but not with the number of qubits $n$. Our numerical simulations focus on frequency spaces increasing polynomially with $n$. However, it is possible to construct data-generators with exponentially many equally probable different accessible frequencies~\cite{shin2023exponential}. In this scenario,~\Cref{th.conv_spectrum} still holds, leading to a Gaussian profile of frequencies with variance $\sigma \in \Theta(2^n \sqrt L)$. Note that exponentially many frequencies require exponentially many tunable parameters to match the number of degrees of freedom. Therefore, data-encoding generators with $\sim e^n$ different frequencies can only aim to efficiently learn functions with sparse Fourier representations, i.e. with only $\mathcal O({\rm poly}(n))$ non-zero Fourier coefficients in a $\mathcal O(e^n)$ frequency space. 

The expressivity results from~\Cref{sec.expressivity} imply direct limitations in the attainable hypothesis functions, but also give an intuition on how to amplify high-frequency Fourier components, or in other words how to maximize the Lipschitz constant. The only recipe to obtain a high-frequency Fourier profile is by repeatedly amplifying the eigenvectors corresponding to extreme eigenvalues of the data-encoding generator. This yields an extremal case as far as possible from the average case where unitaries are sampled from the Haar distribution. Without loss of generality, we may choose the ground state. The first step of the circuit would have to transform the initial state into the ground state of the data-encoding generator. For $k$-local Hamiltonians with $k \geq 2$, this problem is QMA-complete~\cite{kempe2005complexity}, and finding the hypothesis function with maximum high-frequency content implies repeatedly solving a QMA-complete problem. An example is choosing the data generator to be the Hamiltonian of a transverse-field Ising model, constructed on an arbitrary graph. Such PQC does not suffer from BPs if the parameters respect the permutation invariance of the graph~\cite{larocca2021theory}. Therefore, reaching the hypothesis function with maximized high-frequency content is in general hard.  A notable exception appears in layered circuits with one $g$, and $W_i = I$, where setting all parameters $\theta=0$ suffices to maintain the quantum state aligned with the ground state of $g$. Notably, maximizing the Lipschitz constant in the experiments from~\Cref{sec.num_expressivity} is feasible.

We have seen that hypothesis functions produced by layered QRU models have naturally vanishing high-frequency components, therefore limiting their Lipschitz constant. Regularization of the Lipschitz constant yields increased generalization and robustness of classical Neural Networks \cite{gouk2021regularisation, bartlett2017spectrallynormalized}. 
As a consequence, this could hint toward a better generalization capacity of QRU models, in agreement with existing literature~\cite{peters2022generalization}.

The scope of this work is the average behavior of QRU models. It shows concentration properties, similar to other existing results~\cite{fontana2023classical, rudolph2023classical, mcclean2018barren}, and provides useful insights on the internal working principles of the model. This interpretability will be useful to develop QML models with specific properties. It will be possible to investigate protection against dequantization through peaked generalized trigonometric polynomials, in alignment with peaked circuits~\cite{bravyi2023classical}. Restrictions of generators and parameterized steps in the circuits can be applied to constraint the behavior of output models, allowing for systematic exploration of ingredients in QRU. 

\section{Conclusions}\label{sec.conclusions}

We have explored the features of QRU models to understand the implications of injecting data into the better-studied PQCs models. Two main features are studied, first the magnitude of the gradients of the loss function, and second the frequency profile of the hypothesis function output by QRU models. Results were proven analytically and extended to more practical scenarios numerically.

We give analytical bounds for the connection between the variance of gradients in the hypothesis functions of QRU models and the cost functions on corresponding PQCs. Vanishing gradients of hypothesis functions imply vanishing gradients for any cost function to train ML models with, thus preventing trainability. The difference between QRU models and PQCs can be quantified by measuring the effect of adding data to the circuit, averaging over the parameter space. This is quantified by the coined concept of {\sl absorption witness}. If data can be re-absorbed in a shift of parameters, then the gradients for PQCs and QRU models take similar value ranges. Results can be further simplified for the case of layered ansatzes. These results provide insights into the construction of QRU models protected against the phenomenon of BP, by using existing knowledge on PQC that do not exhibit BPs.

We prove also that QRU models suffer from vanishing high frequencies.  Each additional data encoding operation corresponds to an additional convolution of the current spectrum with the data generator spectrum. As the number of layers, denoted as $L$, becomes large the span of attainable frequencies grows as $\bigO{L}$. However, the central limit theorem dictates that the frequency profile follows a Gaussian distribution, spreading out proportionally to $\sqrt{L}$. Therefore, in practice, only frequencies (approximately) bounded by $\sqrt L$ are available, with the contribution of higher frequencies exponentially vanishing. The vanishing high frequencies have direct consequences on the class of functions attainable by the QRU models. The average of the optimal Lipschitz constant scales with $\sqrt L$, exhibiting an exponentially decaying probability of exceeding this value. These findings offer insights into the inherent limitations of expressivity in QRU models and provide tools for estimating the computational resources required to represent specific datasets effectively.

The results derived in this work broaden our understanding of the properties of QRU models and provide guidelines for the design of re-uploading schemes. As an example, the concept of absorption witness can be employed to select generators ensuring an ansatz with the necessary characteristics to be trainable. For expressivity, adjusting the depth of the model can strike a balance between capturing intricate details in the data and avoiding overfitting. Consequently, we anticipate that these tools and insights will contribute to enhancing the applicability and performance of QRU models. \\

\acknowledgments

The authors thank Elies Gil-Fuster, Vedran Dunjko, Patrick Emonts, and Xavier Bonet-Monroig for their useful comments on this manuscript. This work was supported by CERN through the CERN Quantum Technology Initiative. This work was carried out as part of the quantum computing for earth observation (QC4EO) initiative of ESA $\phi$-lab, partially funded under contract 4000135723/21/I-DT-lr, in the FutureEO program. This work was supported by the Dutch National Growth Fund (NGF), as part of the Quantum Delta NL programme.

\bibliographystyle{quantum}
\bibliography{references, references_extra}

\begin{thebibliography}{10}

\bibitem{preskill2018quantum}
John Preskill.
\newblock ``Quantum {{Computing}} in the {{NISQ}} era and beyond''.
\newblock \href{https://dx.doi.org/10.22331/q-2018-08-06-79}{Quantum {\bf 2},
  79}~(2018).

\bibitem{bharti2022noisy}
Kishor Bharti, Alba {Cervera-Lierta}, Thi~Ha Kyaw, Tobias Haug, Sumner
  {Alperin-Lea}, Abhinav Anand, Matthias Degroote, Hermanni Heimonen, Jakob~S.
  Kottmann, Tim Menke, Wai-Keong Mok, Sukin Sim, Leong-Chuan Kwek, and Al{\'a}n
  {Aspuru-Guzik}.
\newblock ``Noisy intermediate-scale quantum algorithms''.
\newblock \href{https://dx.doi.org/10.1103/RevModPhys.94.015004}{Reviews of
  Modern Physics {\bf 94}, 015004}~(2022).

\bibitem{cerezo2021variational}
M.~Cerezo, Andrew Arrasmith, Ryan Babbush, Simon~C. Benjamin, Suguru Endo,
  Keisuke Fujii, Jarrod~R. McClean, Kosuke Mitarai, Xiao Yuan, Lukasz Cincio,
  and Patrick~J. Coles.
\newblock ``Variational quantum algorithms''.
\newblock \href{https://dx.doi.org/10.1038/s42254-021-00348-9}{Nature Reviews
  Physics {\bf 3}, 625--644}~(2021).

\bibitem{mcclean2021low}
Jarrod~R. McClean, Matthew~P. Harrigan, Masoud Mohseni, Nicholas~C. Rubin,
  Zhang Jiang, Sergio Boixo, Vadim~N. Smelyanskiy, Ryan Babbush, and Hartmut
  Neven.
\newblock ``Low depth mechanisms for quantum optimization''.
\newblock \href{https://dx.doi.org/10.1103/PRXQuantum.2.030312}{PRX Quantum
  {\bf 2}, 030312}~(2021).
\newblock  \href{http://arxiv.org/abs/2008.08615}{arXiv:2008.08615}.

\bibitem{bittel2021training}
Lennart Bittel and Martin Kliesch.
\newblock ``Training variational quantum algorithms is {{NP-hard}}''.
\newblock \href{https://dx.doi.org/10.1103/PhysRevLett.127.120502}{Physical
  Review Letters {\bf 127}, 120502}~(2021).
\newblock  \href{http://arxiv.org/abs/2101.07267}{arXiv:2101.07267}.

\bibitem{peruzzo2014variational}
Alberto Peruzzo, Jarrod McClean, Peter Shadbolt, Man-Hong Yung, Xiao-Qi Zhou,
  Peter~J. Love, Al{\'a}n {Aspuru-Guzik}, and Jeremy~L. O'Brien.
\newblock ``A variational eigenvalue solver on a photonic quantum processor''.
\newblock \href{https://dx.doi.org/10.1038/ncomms5213}{Nature Communications
  {\bf 5}, 4213}~(2014).

\bibitem{mcclean2016theory}
Jarrod~R. McClean, Jonathan Romero, Ryan Babbush, and Al{\'a}n {Aspuru-Guzik}.
\newblock ``The theory of variational hybrid quantum-classical algorithms''.
\newblock \href{https://dx.doi.org/10.1088/1367-2630/18/2/023023}{New Journal
  of Physics {\bf 18}, 023023}~(2016).

\bibitem{ryabinkin2019constrained}
Ilya~G. Ryabinkin, Scott~N. Genin, and Artur~F. Izmaylov.
\newblock ``Constrained {{Variational Quantum Eigensolver}}: {{Quantum Computer
  Search Engine}} in the {{Fock Space}}''.
\newblock \href{https://dx.doi.org/10.1021/acs.jctc.8b00943}{Journal of
  Chemical Theory and Computation {\bf 15}, 249--255}~(2019).

\bibitem{farhi2014quantum}
Edward Farhi, Jeffrey Goldstone, and Sam Gutmann.
\newblock ``A {{Quantum Approximate Optimization Algorithm}}''~(2014).
\newblock  \href{http://arxiv.org/abs/1411.4028}{arXiv:1411.4028}.

\bibitem{cao2019quantum}
Yudong Cao, Jonathan Romero, Jonathan~P. Olson, Matthias Degroote, Peter~D.
  Johnson, M{\'a}ria Kieferov{\'a}, Ian~D. Kivlichan, Tim Menke, Borja
  Peropadre, Nicolas P.~D. Sawaya, Sukin Sim, Libor Veis, and Al{\'a}n
  {Aspuru-Guzik}.
\newblock ``Quantum {{Chemistry}} in the {{Age}} of {{Quantum Computing}}''.
\newblock \href{https://dx.doi.org/10.1021/acs.chemrev.8b00803}{Chemical
  Reviews {\bf 119}, 10856--10915}~(2019).

\bibitem{li2017efficient}
Ying Li and Simon~C. Benjamin.
\newblock ``Efficient {{Variational Quantum Simulator Incorporating Active
  Error Minimization}}''.
\newblock \href{https://dx.doi.org/10.1103/PhysRevX.7.021050}{Physical Review X
  {\bf 7}, 021050}~(2017).

\bibitem{cirstoiu2020variational}
Cristina Cirstoiu, Zoe Holmes, Joseph Iosue, Lukasz Cincio, Patrick~J. Coles,
  and Andrew Sornborger.
\newblock ``Variational {{Fast Forwarding}} for {{Quantum Simulation Beyond}}
  the {{Coherence Time}}''.
\newblock \href{https://dx.doi.org/10.1038/s41534-020-00302-0}{npj Quantum
  Information {\bf 6}, 82}~(2020).
\newblock  \href{http://arxiv.org/abs/1910.04292}{arXiv:1910.04292}.

\bibitem{bharti2021quantumassisted}
Kishor Bharti and Tobias Haug.
\newblock ``Quantum-assisted simulator''.
\newblock \href{https://dx.doi.org/10.1103/PhysRevA.104.042418}{Physical Review
  A {\bf 104}, 042418}~(2021).

\bibitem{mcardle2019variational}
Sam McArdle, Tyson Jones, Suguru Endo, Ying Li, Simon~C. Benjamin, and Xiao
  Yuan.
\newblock ``Variational ansatz-based quantum simulation of imaginary time
  evolution''.
\newblock \href{https://dx.doi.org/10.1038/s41534-019-0187-2}{npj Quantum
  Information {\bf 5}, 1--6}~(2019).

\bibitem{yuan2019theory}
Xiao Yuan, Suguru Endo, Qi~Zhao, Ying Li, and Simon~C. Benjamin.
\newblock ``Theory of variational quantum simulation''.
\newblock \href{https://dx.doi.org/10.22331/q-2019-10-07-191}{Quantum {\bf 3},
  191}~(2019).

\bibitem{mitarai2018quantum}
Kosuke Mitarai, Makoto Negoro, Masahiro Kitagawa, and Keisuke Fujii.
\newblock ``Quantum {{Circuit Learning}}''.
\newblock \href{https://dx.doi.org/10.1103/PhysRevA.98.032309}{Physical Review
  A {\bf 98}, 032309}~(2018).
\newblock  \href{http://arxiv.org/abs/1803.00745}{arXiv:1803.00745}.

\bibitem{schuld2020circuitcentric}
Maria Schuld, Alex Bocharov, Krysta Svore, and Nathan Wiebe.
\newblock ``Circuit-centric quantum classifiers''.
\newblock \href{https://dx.doi.org/10.1103/PhysRevA.101.032308}{Physical Review
  A {\bf 101}, 032308}~(2020).
\newblock  \href{http://arxiv.org/abs/1804.00633}{arXiv:1804.00633}.

\bibitem{havlicek2019supervised}
Vojt{\v e}ch Havl{\'i}{\v c}ek, Antonio~D. C{\'o}rcoles, Kristan Temme, Aram~W.
  Harrow, Abhinav Kandala, Jerry~M. Chow, and Jay~M. Gambetta.
\newblock ``Supervised learning with quantum-enhanced feature spaces''.
\newblock \href{https://dx.doi.org/10.1038/s41586-019-0980-2}{Nature {\bf 567},
  209--212}~(2019).

\bibitem{schuld2021supervised}
Maria Schuld.
\newblock ``Supervised quantum machine learning models are kernel
  methods''~(2021).
\newblock  \href{http://arxiv.org/abs/2101.11020}{arXiv:2101.11020}.

\bibitem{otterbach2017unsupervised}
J.~S. Otterbach, R.~Manenti, N.~Alidoust, A.~Bestwick, M.~Block, B.~Bloom,
  S.~Caldwell, N.~Didier, E.~Schuyler Fried, S.~Hong, P.~Karalekas, C.~B.
  Osborn, A.~Papageorge, E.~C. Peterson, G.~Prawiroatmodjo, N.~Rubin, Colm~A.
  Ryan, D.~Scarabelli, M.~Scheer, E.~A. Sete, P.~Sivarajah, Robert~S. Smith,
  A.~Staley, N.~Tezak, W.~J. Zeng, A.~Hudson, Blake~R. Johnson, M.~Reagor,
  M.~P. {da Silva}, and C.~Rigetti.
\newblock ``Unsupervised {{Machine Learning}} on a {{Hybrid Quantum
  Computer}}''~(2017).
\newblock  \href{http://arxiv.org/abs/1712.05771}{arXiv:1712.05771}.

\bibitem{zoufal2019quantum}
Christa Zoufal, Aur{\'e}lien Lucchi, and Stefan Woerner.
\newblock ``Quantum {{Generative Adversarial Networks}} for learning and
  loading random distributions''.
\newblock \href{https://dx.doi.org/10.1038/s41534-019-0223-2}{npj Quantum
  Information {\bf 5}, 1--9}~(2019).

\bibitem{zoufal2021variational}
Christa Zoufal, Aur{\'e}lien Lucchi, and Stefan Woerner.
\newblock ``Variational quantum {{Boltzmann}} machines''.
\newblock \href{https://dx.doi.org/10.1007/s42484-020-00033-7}{Quantum Machine
  Intelligence {\bf 3}, 7}~(2021).

\bibitem{dallaire-demers2018quantum}
Pierre-Luc {Dallaire-Demers} and Nathan Killoran.
\newblock ``Quantum generative adversarial networks''.
\newblock \href{https://dx.doi.org/10.1103/PhysRevA.98.012324}{Physical Review
  A {\bf 98}, 012324}~(2018).
\newblock  \href{http://arxiv.org/abs/1804.08641}{arXiv:1804.08641}.

\bibitem{schuld2019quantum}
Maria Schuld and Nathan Killoran.
\newblock ``Quantum machine learning in feature {{Hilbert}} spaces''.
\newblock \href{https://dx.doi.org/10.1103/PhysRevLett.122.040504}{Physical
  Review Letters {\bf 122}, 040504}~(2019).
\newblock  \href{http://arxiv.org/abs/1803.07128}{arXiv:1803.07128}.

\bibitem{vidal2020input}
Javier~Gil Vidal and Dirk~Oliver Theis.
\newblock ``Input {{Redundancy}} for {{Parameterized Quantum
  Circuits}}''~(2020).
\newblock  \href{http://arxiv.org/abs/1901.11434}{arXiv:1901.11434}.

\bibitem{schuld2021effect}
Maria Schuld, Ryan Sweke, and Johannes~Jakob Meyer.
\newblock ``The effect of data encoding on the expressive power of variational
  quantum machine learning models''.
\newblock \href{https://dx.doi.org/10.1103/PhysRevA.103.032430}{Physical Review
  A {\bf 103}, 032430}~(2021).
\newblock  \href{http://arxiv.org/abs/2008.08605}{arXiv:2008.08605}.

\bibitem{perez-salinas2020data}
Adri{\'a}n {P{\'e}rez-Salinas}, Alba {Cervera-Lierta}, Elies {Gil-Fuster}, and
  Jos{\'e}~I. Latorre.
\newblock ``Data re-uploading for a universal quantum classifier''.
\newblock \href{https://dx.doi.org/10.22331/q-2020-02-06-226}{Quantum {\bf 4},
  226}~(2020).

\bibitem{perez-salinas2021determining}
Adri{\'a}n {P{\'e}rez-Salinas}, Juan {Cruz-Martinez}, Abdulla~A. Alhajri, and
  Stefano Carrazza.
\newblock ``Determining the proton content with a quantum computer''.
\newblock \href{https://dx.doi.org/10.1103/PhysRevD.103.034027}{Physical Review
  D {\bf 103}, 034027}~(2021).

\bibitem{sim2019expressibility}
Sukin Sim, Peter~D. Johnson, and Al{\'a}n {Aspuru-Guzik}.
\newblock ``Expressibility and {{Entangling Capability}} of {{Parameterized
  Quantum Circuits}} for {{Hybrid Quantum-Classical Algorithms}}''.
\newblock \href{https://dx.doi.org/10.1002/qute.201900070}{Advanced Quantum
  Technologies {\bf 2}, 1900070}~(2019).

\bibitem{cybenko1989approximation}
G.~Cybenko.
\newblock ``Approximation by superpositions of a sigmoidal function''.
\newblock \href{https://dx.doi.org/10.1007/BF02551274}{Mathematics of Control,
  Signals and Systems {\bf 2}, 303--314}~(1989).

\bibitem{hornik1991approximation}
Kurt Hornik.
\newblock ``Approximation capabilities of multilayer feedforward networks''.
\newblock \href{https://dx.doi.org/10.1016/0893-6080(91)90009-T}{Neural
  Networks {\bf 4}, 251--257}~(1991).

\bibitem{perez-salinas2021one}
Adri{\'a}n {P{\'e}rez-Salinas}, David {L{\'o}pez-N{\'u}{\~n}ez}, Artur
  {Garc{\'i}a-S{\'a}ez}, P.~{Forn-D{\'i}az}, and Jos{\'e}~I. Latorre.
\newblock ``One qubit as a universal approximant''.
\newblock \href{https://dx.doi.org/10.1103/PhysRevA.104.012405}{Physical Review
  A {\bf 104}, 012405}~(2021).

\bibitem{anschuetz2022barren}
Eric~R. Anschuetz and Bobak~T. Kiani.
\newblock ``Beyond {{Barren Plateaus}}: {{Quantum Variational Algorithms Are
  Swamped With Traps}}''~(2022).
\newblock  \href{http://arxiv.org/abs/2205.05786}{arXiv:2205.05786}.

\bibitem{mcclean2018barren}
Jarrod~R. McClean, Sergio Boixo, Vadim~N. Smelyanskiy, Ryan Babbush, and
  Hartmut Neven.
\newblock ``Barren plateaus in quantum neural network training landscapes''.
\newblock \href{https://dx.doi.org/10.1038/s41467-018-07090-4}{Nature
  Communications {\bf 9}, 4812}~(2018).

\bibitem{thanasilp2023subtleties}
Supanut Thanasilp, Samson Wang, Nhat~A. Nghiem, Patrick~J. Coles, and
  M.~Cerezo.
\newblock ``Subtleties in the trainability of quantum machine learning
  models''.
\newblock \href{https://dx.doi.org/10.1007/s42484-023-00103-6}{Quantum Machine
  Intelligence {\bf 5}, 21}~(2023).
\newblock  \href{http://arxiv.org/abs/2110.14753}{arXiv:2110.14753}.

\bibitem{holmes2022connecting}
Zo{\"e} Holmes, Kunal Sharma, M.~Cerezo, and Patrick~J. Coles.
\newblock ``Connecting {{Ansatz Expressibility}} to {{Gradient Magnitudes}} and
  {{Barren Plateaus}}''.
\newblock \href{https://dx.doi.org/10.1103/PRXQuantum.3.010313}{PRX Quantum
  {\bf 3}, 010313}~(2022).

\bibitem{hubregtsen2021evaluation}
Thomas Hubregtsen, Josef Pichlmeier, Patrick Stecher, and Koen Bertels.
\newblock ``Evaluation of parameterized quantum circuits: On the relation
  between classification accuracy, expressibility, and entangling capability''.
\newblock \href{https://dx.doi.org/10.1007/s42484-021-00038-w}{Quantum Machine
  Intelligence {\bf 3}, 9}~(2021).

\bibitem{larocca2022diagnosing}
Martin Larocca, Piotr Czarnik, Kunal Sharma, Gopikrishnan Muraleedharan,
  Patrick~J. Coles, and M.~Cerezo.
\newblock ``Diagnosing {{Barren Plateaus}} with {{Tools}} from {{Quantum
  Optimal Control}}''.
\newblock \href{https://dx.doi.org/10.22331/q-2022-09-29-824}{Quantum {\bf 6},
  824}~(2022).
\newblock  \href{http://arxiv.org/abs/2105.14377}{arXiv:2105.14377}.

\bibitem{larocca2021theory}
Martin Larocca, Nathan Ju, Diego {Garc{\'i}a-Mart{\'i}n}, Patrick~J. Coles, and
  M.~Cerezo.
\newblock ``Theory of overparametrization in quantum neural networks''~(2021).
\newblock  \href{http://arxiv.org/abs/2109.11676}{arXiv:2109.11676}.

\bibitem{caro2021encodingdependent}
Matthias~C. Caro, Elies {Gil-Fuster}, Johannes~Jakob Meyer, Jens Eisert, and
  Ryan Sweke.
\newblock ``Encoding-dependent generalization bounds for parametrized quantum
  circuits''.
\newblock \href{https://dx.doi.org/10.22331/q-2021-11-17-582}{Quantum {\bf 5},
  582}~(2021).

\bibitem{jerbi2023quantum}
Sofiene Jerbi, Lukas~J. Fiderer, Hendrik Poulsen~Nautrup, Jonas~M. K{\"u}bler,
  Hans~J. Briegel, and Vedran Dunjko.
\newblock ``Quantum machine learning beyond kernel methods''.
\newblock \href{https://dx.doi.org/10.1038/s41467-023-36159-y}{Nature
  Communications {\bf 14}, 517}~(2023).

\bibitem{munoz2015exploratory}
Mario~A. Mu{\~n}oz, Michael Kirley, and Saman~K. Halgamuge.
\newblock ``Exploratory {{Landscape Analysis}} of {{Continuous Space
  Optimization Problems Using Information Content}}''.
\newblock \href{https://dx.doi.org/10.1109/TEVC.2014.2302006}{IEEE Transactions
  on Evolutionary Computation {\bf 19}, 74--87}~(2015).

\bibitem{pascanu2013difficulty}
Razvan Pascanu, Tomas Mikolov, and Yoshua Bengio.
\newblock ``On the difficulty of training {{Recurrent Neural
  Networks}}''~(2013).
\newblock  \href{http://arxiv.org/abs/1211.5063}{arXiv:1211.5063}.

\bibitem{friedrich2022escaping}
Tobias Friedrich, Timo K{\"o}tzing, Martin~S. Krejca, and Amirhossein Rajabi.
\newblock ``Escaping {{Local Optima}} with~{{Local Search}}: {{A Theory-Driven
  Discussion}}''.
\newblock In G{\"u}nter Rudolph, Anna~V. Kononova, Hern{\'a}n Aguirre, Pascal
  Kerschke, Gabriela Ochoa, and Tea Tu{\v s}ar, editors, Parallel {{Problem
  Solving}} from {{Nature}} -- {{PPSN XVII}}.
\newblock \href{https://dx.doi.org/10.1007/978-3-031-14721-0_31}{Pages
  442--455}.
\newblock Lecture {{Notes}} in {{Computer Science}}Cham~(2022). Springer
  International Publishing.

\bibitem{raghu2017expressive}
Maithra Raghu, Ben Poole, Jon Kleinberg, Surya Ganguli, and Jascha
  {Sohl-Dickstein}.
\newblock ``On the {{Expressive Power}} of {{Deep Neural Networks}}''~(2017).
\newblock  \href{http://arxiv.org/abs/1606.05336}{arXiv:1606.05336}.

\bibitem{schatzki2022theoretical}
Louis Schatzki, Martin Larocca, Quynh~T. Nguyen, Frederic Sauvage, and
  M.~Cerezo.
\newblock ``Theoretical {{Guarantees}} for {{Permutation-Equivariant Quantum
  Neural Networks}}''~(2022).
\newblock  \href{http://arxiv.org/abs/2210.09974}{arXiv:2210.09974}.

\bibitem{cerezo2021cost}
M.~Cerezo, Akira Sone, Tyler Volkoff, Lukasz Cincio, and Patrick~J. Coles.
\newblock ``Cost function dependent barren plateaus in shallow parametrized
  quantum circuits''.
\newblock \href{https://dx.doi.org/10.1038/s41467-021-21728-w}{Nature
  Communications {\bf 12}, 1791}~(2021).

\bibitem{barenco1995elementary}
Adriano Barenco, Charles~H. Bennett, Richard Cleve, David~P. DiVincenzo, Norman
  Margolus, Peter Shor, Tycho Sleator, John~A. Smolin, and Harald Weinfurter.
\newblock ``Elementary gates for quantum computation''.
\newblock \href{https://dx.doi.org/10.1103/PhysRevA.52.3457}{Physical Review A
  {\bf 52}, 3457--3467}~(1995).

\bibitem{github}
Alice Barthe and Adrián Pérez-Salinas.
\newblock ``Github repository: \texttt{QRU\_average}''~(2023).

\bibitem{perez-salinas2023analyzing}
Adri{\'a}n {P{\'e}rez-Salinas}, Hao Wang, and Xavier {Bonet-Monroig}.
\newblock ``Analyzing variational quantum landscapes with information
  content''~(2023).
\newblock  \href{http://arxiv.org/abs/2303.16893}{arXiv:2303.16893}.

\bibitem{olkin1964multivariate}
Ingram Olkin and Herman Rubin.
\newblock ``Multivariate {{Beta Distributions}} and {{Independence Properties}}
  of the {{Wishart Distribution}}''.
\newblock \href{https://dx.doi.org/10.1214/aoms/1177703748}{The Annals of
  Mathematical Statistics {\bf 35}, 261--269}~(1964).

\bibitem{billingsley1995probability}
Patrick Billingsley.
\newblock ``Probability and {{Measure}}''.
\newblock Wiley. ~(1995).

\bibitem{schreiber2023classical}
Franz~J. Schreiber, Jens Eisert, and Johannes~Jakob Meyer.
\newblock ``Classical surrogates for quantum learning models''.
\newblock \href{https://dx.doi.org/10.1103/PhysRevLett.131.100803}{Physical
  Review Letters {\bf 131}, 100803}~(2023).
\newblock  \href{http://arxiv.org/abs/2206.11740}{arXiv:2206.11740}.

\bibitem{shin2023exponential}
S.~Shin, Y.~S. Teo, and H.~Jeong.
\newblock ``Exponential data encoding for quantum supervised learning''.
\newblock \href{https://dx.doi.org/10.1103/PhysRevA.107.012422}{Physical Review
  A {\bf 107}, 012422}~(2023).

\bibitem{kempe2005complexity}
Julia Kempe, Alexei Kitaev, and Oded Regev.
\newblock ``The {{Complexity}} of the {{Local Hamiltonian Problem}}''~(2005).
\newblock
  \href{http://arxiv.org/abs/quant-ph/0406180}{arXiv:quant-ph/0406180}.

\bibitem{gouk2021regularisation}
Henry Gouk, Eibe Frank, Bernhard Pfahringer, and Michael~J. Cree.
\newblock ``Regularisation of neural networks by enforcing {{Lipschitz}}
  continuity''.
\newblock \href{https://dx.doi.org/10.1007/s10994-020-05929-w}{Machine Learning
  {\bf 110}, 393--416}~(2021).

\bibitem{bartlett2017spectrallynormalized}
Peter Bartlett, Dylan~J. Foster, and Matus Telgarsky.
\newblock ``Spectrally-normalized margin bounds for neural networks''~(2017).
\newblock  \href{http://arxiv.org/abs/1706.08498}{arXiv:1706.08498}.

\bibitem{peters2022generalization}
Evan Peters and Maria Schuld.
\newblock ``Generalization despite overfitting in quantum machine learning
  models''~(2022).
\newblock  \href{http://arxiv.org/abs/2209.05523}{arXiv:2209.05523}.

\bibitem{fontana2023classical}
Enrico Fontana, Manuel~S. Rudolph, Ross Duncan, Ivan Rungger, and Cristina
  C{\^i}rstoiu.
\newblock ``Classical simulations of noisy variational quantum
  circuits''~(2023).
\newblock  \href{http://arxiv.org/abs/2306.05400}{arXiv:2306.05400}.

\bibitem{rudolph2023classical}
Manuel~S. Rudolph, Enrico Fontana, Zo{\"e} Holmes, and Lukasz Cincio.
\newblock ``Classical surrogate simulation of quantum systems with
  {{LOWESA}}''~(2023).
\newblock  \href{http://arxiv.org/abs/2308.09109}{arXiv:2308.09109}.

\bibitem{bravyi2023classical}
Sergey Bravyi, David Gosset, and Yinchen Liu.
\newblock ``Classical simulation of peaked shallow quantum circuits''~(2023).
\newblock  \href{http://arxiv.org/abs/2309.08405}{arXiv:2309.08405}.

\bibitem{boixo2018characterizing}
Sergio Boixo, Sergei~V. Isakov, Vadim~N. Smelyanskiy, Ryan Babbush, Nan Ding,
  Zhang Jiang, Michael~J. Bremner, John~M. Martinis, and Hartmut Neven.
\newblock ``Characterizing {{Quantum Supremacy}} in {{Near-Term Devices}}''.
\newblock \href{https://dx.doi.org/10.1038/s41567-018-0124-x}{Nature Physics
  {\bf 14}, 595--600}~(2018).
\newblock  \href{http://arxiv.org/abs/1608.00263}{arXiv:1608.00263}.

\bibitem{bailey1992distributional}
Ralph~W. Bailey.
\newblock ``Distributional {{Identities}} of {{Beta}} and {{Chi-Squared
  Variates}}: {{A Geometrical Interpretation}}''.
\newblock \href{https://dx.doi.org/10.2307/2684178}{The American Statistician
  {\bf 46}, 117--120}~(1992).
\newblock  \href{http://arxiv.org/abs/2684178}{arXiv:2684178}.

\bibitem{hoeffding1963probability}
Wassily Hoeffding.
\newblock ``Probability {{Inequalities}} for {{Sums}} of {{Bounded Random
  Variables}}''.
\newblock \href{https://dx.doi.org/10.1080/01621459.1963.10500830}{Journal of
  the American Statistical Association {\bf 58}, 13--30}~(1963).

\bibitem{zee2010quantum}
A.~Zee.
\newblock ``Quantum field theory in a nutshell''.
\newblock In a Nutshell. Princeton University Press. Princeton, N.J~(2010).
\newblock 2nd ed edition.

\bibitem{user268722012answer}
{user26872}.
\newblock ``Answer to "reference for multidimensional gaussian
  integral"''~(2012).

\end{thebibliography}

\newpage
\onecolumngrid

\appendix

\section{Proofs}

\subsection{Proof of~\Cref{th.var_bounds}}\label{app.var_bounds}
We begin by explicitly writing the derivatives of the hypothesis function
    \begin{equation}\label{eq.trace_form}
        \partial_j h_{\btheta}(x) = \Tr{U_{R, j}(\btheta_{R, j}, x) \ \rho_0 U^\dagger_{R, j}(\btheta_{R, j}, x) \left[ V_j, U^\dagger_{L, j}(\btheta_{L, j}, x) H U_{L, j}(\btheta_{L, j}, x) \right]}.
    \end{equation}
    In this equation, the indices $R, L$ indicate all operations before and after the $j-$th operation. We redefine the quantities for readability
    \begin{align}
        \rho_j(\btheta_{R, j}, x) & = U_{R, j}(\btheta_{R, j}, x) \ \rho_0 \ U^\dagger_{R, j}(\btheta_{R, j}, x) \\
        H_j(\btheta_{L, j}, x) & = U^\dagger_{L, j}(\btheta_{L, j}, x) \ H \ U_{L, j}(\btheta_{L, j}, x)
    \end{align}

    The variance of these derivatives over $\Btheta$ is given by 
    \begin{equation}\label{eq.variance_def}
        \Ex{\Var[\Btheta]{\partial_j h_{\btheta}(x)}} = \E[\Btheta]{\Var[X]{\partial_j h_{\btheta}(x)}}
        = \Er{\El{\Ex{\left( \partial_j h_{\btheta}(x)\right)^2}}},
    \end{equation}
    where we assume no correlation between the parameters in the left and right parts of the circuit.
    By calling the property $\Tr{A\otimes B} = \Tr A \Tr B$, we can plug~\Cref{eq.trace_form} into~\Cref{eq.variance_def} to obtain
    \begin{equation}\label{eq.integral_full}
        \Var[\Btheta, X]{\partial_j h_{\btheta}(x)} = 
        \Er{\El{\Ex{ \Tr{\rho_{j}(\btheta_{R, j}, x)^{\otimes 2}\left[V_j, H_{j}(\btheta_{L, j}, x)\right]^{\otimes 2}}}}} 
    \end{equation}

    We aim to describe this quantity in terms of the difference between the QML models, partially described by data $x$, and their corresponding PQC models, where $x = 0$. We define the corresponding difference operators 
    \begin{align}
    B_{R, j}^{(t)}(\btheta_{R, j}, x; \rho_0) & = \rho_j^{\otimes t}(\btheta_{R, j}, x) - \rho_j^{\otimes t}(\btheta_{R, j}, 0) \\ 
    B_{L, j}^{(t)}(\btheta_{L, j}, x; H)      & =    H_j^{\otimes t}(\btheta_{L, j}, x) -   H_j^{\otimes t}(\btheta_{L, j}, 0).
    \end{align}
    We rearrange the terms in the integrand of~\Cref{eq.integral_full} as
    \begin{align}
        \Tr{\rho_{j}(\btheta_{R, j}, x)^{\otimes 2}\left[V_j, H_{j}(\btheta_{L, j}, x)\right]^{\otimes 2}} & =  \label{eq.variance}\\
        \Tr{\rho_j(\btheta_{R, j}, 0)^{\otimes 2}[V_j, H_j(\btheta_{R, j}, x)]^{\otimes 2}} & + \Tr{B_{R, j}^{(2)}(\btheta_{R, j}, x;\rho_0)[V_j, H_j(\btheta_{L, j}, x)]^{\otimes 2}} = \\ 
        \Tr{H_j(\btheta_{R, j}, x)^{\otimes 2}[\rho_j(\btheta_{R, j}, 0), V_j]^{\otimes 2}} & + \Tr{B_{R, j}^{(2)}(\btheta_{R, j}, x;\rho_0)[V_j, H_j(\btheta_{L, j}, x)]^{\otimes 2}} = \\
        & \Tr{\rho_j(\btheta_{R, j}, 0)^{\otimes 2}[V_j, H_j(\btheta_{R, j}, 0)]^{\otimes 2}} + \label{eq.var_pqc}\\
        & \Tr{B_{R, j}^{(2)}(\btheta_{R, j}, x;\rho_0)[V_j, H_j(\btheta_{L, j}, x)]^{\otimes 2}} +\label{eq.var_rho}\\  
        & \Tr{B_{L, j}^{(2)}(\btheta_{L, j}, x; H)[\rho_j(\btheta_{R, j}, 0), V_j]^{\otimes 2}}  \label{eq.var_h}
    \end{align}
    by recalling the identities $\Tr{A^{\otimes 2}} = \Tr{A}^2$ and $\Tr{A[B, C]} =\Tr{B[C, A]} =\Tr{C[A, B]}$.
    The term in~\Cref{eq.var_pqc} corresponds to the standard variance in PQC. We denote it simply as $\Var[\Btheta]{\partial_j h_\btheta(0)}.$

    We move our attention now~\Cref{eq.var_rho}. This term measures the difference between QRU models and PQC in the right part of the quantum circuit. Assuming that the right and left parameters are uncorrelated, we can rewrite
   
    \begin{multline}\label{eq.uncorrelation}
    \Er{\El{\Ex{ \Tr{B_{R, j}^{(2)}(\btheta_{R, j}, x;\rho_0)\left[V_j, H(\btheta_{L, j}, x)\right]^{\otimes 2}}}}} = \\ \Tr{\left(\Ex{\Er{B_{R, j}^{(2)}(\btheta_{R, j}, x;\rho_0)}}\right)\left[V_j, \El{H(\btheta_{L, j}, x)}\right]^{\otimes 2}}
\end{multline}
Using von Neumann's trace and Hölder inequalities, with Schatten norms
 \begin{equation}
        \vert \Tr{A B} \vert \leq \norm{A}_1 \norm{B}_{\infty}, 
    \end{equation}
in~\Cref{eq.uncorrelation} together with the triangular and Cauchy-Schwarz inequality we obtain
\begin{multline}
    \left\vert\Ex{\Tr{\Er{B_{R, j}^{(2)}(\btheta_{R, j}, x;\rho_0)}}\left[V_j, \El{H(\btheta_{L, j}, x)}\right]^{\otimes 2}} \right\vert \leq \\
    \Ex{\left\vert\Tr{\Er{B_{R, j}^{(2)}(\btheta_{R, j}, x;\rho_0)}}\left[V_j, \El{H(\btheta_{L, j}, x)}\right]^{\otimes 2} \right\vert} \leq \\
    \Ex{\norm{\El{\left[V_j, H(\btheta_{L, j}, x)\right]^{\otimes 2}}}_\infty \ \norm{\Er{B_{R, j}(\btheta_{R, j}, x;\rho_0)}}_1}.
\end{multline}
Substituting in the previous equation the property~\cite{holmes2022connecting}
\begin{equation}\label{eq.norm_bound}
    \norm{\left[V_j, H(\btheta_{L, j}, x)\right]^{\otimes 2}}_\infty \leq 4 \norm{V_j}_\infty^2 \norm{H}_\infty^2, 
\end{equation}
and defining
\begin{equation}
    \B^{(2)}_{R, j}(\rho_0) =  \Ex{\norm{\Er B^{(2)}_{R, j}(\btheta_{R, j}, x;\rho_0)}_1},
\end{equation}
we obtain
\begin{equation}\label{eq.bound_rho}
    \Er{\El{\Ex{\Tr{B_{R, j}^{(2)}(\btheta_{R, j}, x;\rho_0)\left[V_j, H(\btheta_{L, j}, x)\right]^{\otimes 2}}}}} \leq 4 \norm{V_j}_\infty^2 \norm{H}_\infty^2 \ \B^{(2)}_{R, j}(\rho_0).
\end{equation}
Following the same steps for~\Cref{eq.var_h}, we can bound this quantity as
\begin{equation}\label{eq.bound_h}
     \Er{\El{\Ex{ \Tr{B_{L, j}^{(2)}(\btheta_{L, j}, x; H)[V_j, \rho_R(\btheta_{R, j}, 0)]^{\otimes 2}}}}} \leq 4 \norm{V_j}_\infty^2 \norm{\rho_0}_\infty^2 \ \B^{(2)}_{L, j}(H),
\end{equation}
where we defined analogously
\begin{equation}
    \B^{(2)}_{L, j}(H) =  \Ex{\norm{ \El{B^{(2)}_{L, j}(\btheta_{L, j}, x;H)}}_1}.
\end{equation}
Notice that we could interchange the $x$-dependency in~\Cref{eq.var_h} and~\Cref{eq.var_rho} with no effect in the final bounds. The reason is that~\Cref{eq.norm_bound} eliminates the $x$-dependency in the term where it is applied. 
We can compact the results from Equations~\eqref{eq.bound_rho} and \eqref{eq.bound_h} using the triangular inequality in
\begin{equation}
    \left\vert\Var[\Btheta]{\partial_j h_\btheta(0)} - \Var[\Btheta]{\Ex{\partial_j h_{\btheta}(x)}}\right\vert \leq 4 \norm{V_j}_\infty^2 \left( \norm{H}_\infty^2\ \B^{(2)}_{R, j}(\rho_0) +\ \norm{\rho_0}_\infty^2\ \B^{(2)}_{L, j}(h) \right)
\end{equation}
\qed

\subsection{Proof of~\Cref{le.layers}}\label{app.layers}
We start by bounding the right absorption witness of the $(l+1)$-th layer with the triangular and Hölder's inequality as
\begin{multline}
    \B^{(2)}_{R, l+1}(\rho_0) = \Ex{\norm{\E[\Btheta_{l + 1}]{u(\btheta_{l+1})^{\otimes 2} V(x)^{\otimes 2}\ \Erl{\rho_l(\btheta_{R, l}, x)}V^\dagger(x)^{\otimes 2}u^\dagger(\btheta_{l+1})^{\otimes 2}}}_2} \leq \\
    \Ex{\norm{\E[\Btheta_{l + 1}]{u(\btheta_{l+1})^{\otimes 2} V(x)^{\otimes 2}\ \Erl{B^{(2)}_{R, l}(\btheta_{R, l}, x;\rho_0)}V^\dagger(x)^{\otimes 2}u^\dagger(\btheta_{l+1})^{\otimes 2}}}_1} + \\ 
    \Ex{\norm{\E[\Btheta_{l + 1}]{u(\btheta_{l+1})^{\otimes 2} V(x)^{\otimes 2} - u(\btheta_{l+1})^{\otimes 2}}}_1 \norm{\Erl{\rho_l(\btheta_{R, l}, 0)}}_\infty}
\end{multline}

The second term can be identified as the layerwise absorption witness from~\Cref{def.absorption_layer}. The first term of the equation above can be bounded as
\begin{multline}
    \Ex{\norm{\E[\Btheta_{l + 1}]{u(\btheta_{l+1})^{\otimes 2} V(x)^{\otimes 2}\ \Erl{B^{(2)}_{R, l}(\btheta_{R, l}, x;\rho_0)}V^\dagger(x)^{\otimes 2}u^\dagger(\btheta_{l+1})^{\otimes 2}}}_1} \leq \\
    \Ex{\E[\Btheta_{l + 1}]{\norm{u(\btheta_{l+1})^{\otimes 2} V(x)^{\otimes 2}\ \Erl{B^{(2)}_{R, l}(\btheta_{R, l}, x;\rho_0)}V^\dagger(x)^{\otimes 2}u^\dagger(\btheta_{l+1})^{\otimes 2}}_1}} = \B^{(2)}_{R, l}(\rho_0).
\end{multline}

Arranging both results together we can find
\begin{equation}
    \B^{(2)}_{R, l+1}(\rho_0) \leq \B^{(2)}_{R, l}(\rho_0) + \norm{\rho_0}_\infty^2 \A_{l+1}^{(2)}.
\end{equation}
Equivalently for the left part of the circuit, and counting layers backward we find
\begin{equation}
    \B^{(2)}_{L, l}(H) \leq \B^{(2)}_{L, l+1}(\rho_0) + \norm{H}_\infty^2 \A_{l}^{(2)}.
\end{equation}
\qed

\subsection{Details on harmonic representation of QRU models}\label{app.harmonic_qru}
As stated in~\Cref{eq.freq_repr}, the wavefunction after a re-uploading circuit and before measurement can be expressed as 
\begin{equation}
    \ket{\psi(x)} = \sum_{j=1}^{2^n} \sum_{k = -K}^K c_{k, j} e^{i k x} \ket{j}.
\end{equation}
The coefficients $c_{k, j}$ form the matrix $\bm C \in \mathbb C^{2^n \times (2K + 1)}$. Each column (indexed with $k$) represents the corresponding term $e^{ikx}$. The states $\ket j$ are elements of any basis of choice. Each row (indexed with $j$) corresponds to the $x$-dependent amplitude attached $\ket j$. The quantity $p_j(x) = \braket{j}{\psi(x)}$ is a trigonometric polynomial,  $p_j(x) = \sum_{k = -K}^K c_{k, j} e^{i k x}$. Such polynomial can be represented as a vector $p_j = \left\{ c_{k, j}\right\}_{-K\leq k \leq +K}$. In this vector representation, the multiplication of polynomials corresponds to convolution as
\begin{equation}
    p(x) q(x) = \sum_{k = -K_p}^{K_p} p_k e^{i k x} \sum_{k = -K_q}^{K_q} q_k e^{i k x} = (p*q)(x),
\end{equation}
We consider the three operations that can be applied to the harmonic representation. 
\paragraph{Parametrized gates} Applying a parameterized gate on the quantum state maps into applying the unitary representation of that gate to each column individually, the same way it would be done in state vector simulation for each (fixed) $k$.
\paragraph{Data-encoding gates} Adding a data-encoding gate involves convolution, when $\boldsymbol{C}$ is expressed in the eigenbasis of the data generator. Each row $j$ corresponds to the $j$-th (integer) eigenvalue $\lambda_j$ from the spectrum of the data-encoding generator,  $\K_g$. The vector representation of polynomials $p_j$ is convoluted with the vector $e_{\lambda_j} = [\delta_{k=\lambda_j}]_{-K\leq k \leq +K}$. 
\paragraph{Measurement} For the measurement, we express $\boldsymbol{C}$ the basis of the observable. Secondly the representation the transpose conjugate of the wavefunction is computed from $\boldsymbol{C}$, by taking its conjugate and reverting the rows indexing $c_{j,k} = c_{j,-k}$. Finally the hypothesis function $h_{\btheta}(x)$ can be obtained as a linear combination of the rows of the result of the convolution weighted by the corresponding eigenvalue of the measurement operator.

\subsection{Proof of~\Cref{th.conv_spectrum}}\label{app.conv_spectrum}
We assume now that the parameterized gates between two consecutive encoding steps are random unitaries sampled from the Haar measure of the group $\SU(N)$. Notice that data-independent operations leave the norm of coefficients associated with the same frequency invariant. Random unitaries output random states for any input. Random states give rise to a probability distribution sampled from a uniform Dirichlet~\cite{olkin1964multivariate}, also known as Porter-Thomas~\cite{boixo2018characterizing} distribution. Under this assumption, no basis has any preference over any other, and the application of the data encoding layer {\sl transports} as many coefficients as dictated by the spectrum $\K_{g_j}$ to the corresponding new frequencies. Notice that it is irrelevant which coefficients are transported, and also they are randomly chosen. The weights for the elements of each frequency are then described by a Dirichlet distribution with parameters given by the convoluted spectrum. Formally, 
\begin{equation}
    \sum_j \left\vert c_{k, j} \right\vert^2 \sim \Dir\left(\left(\K_{g_1} \ast \ldots \ast \K_{g_j} \ast \ldots \ast \K_{g_L}\right)(k)\right),
\end{equation}
where $*$ denotes the convolution operator. Notice that the index $k$ runs over all non-zero entries of the convoluted spectra. \qed

\subsection{Dirichlet distribution}~\label{app.dirichlet}

\begin{definition}[Dirichlet distribution~\cite{olkin1964multivariate}]\label{def.dirichlet}
The Dirichlet distribution $\bm x \sim \Dirichlet$ parameterized by $\alpha\in\mathbb{R}_{>0}^N$ is supported on the $(N-1)$-standard simplex, i.e., $\bm x=(x_1, x_2, \ldots, x_N), \Vert \bm x \Vert_1 = 1$. It has the following probability density function with respect to the Lebesgue measure on $\mathbb{R}^{N-1}$: 
\begin{equation}
    f_{\rm Dir}(\bm x, \bm \alpha) = \frac{\Gamma(\Vert \bm\alpha\Vert_1)}{\prod_{i = 1}^N\Gamma(\alpha_i)}\prod_{i = 1}^N x_i^{\alpha_i - 1}.
\end{equation}
\end{definition}

In this definition, $\Gamma(\cdot)$ is defined as
\begin{equation}
    \Gamma(z) = \int_0^\infty t^{z -1} e^{-t} dt,
\end{equation}
being the complex extension of the factorial for positive integers 
\begin{equation}
    \Gamma(n) = (n - 1)!.
\end{equation}

The Dirichlet distribution admits straightforward analytical calculations for the statistical moments of arbitrary order $\bm k = (k_1, k_2, \ldots, k_N)$,
\begin{equation}\label{eq.dir_moments}
    \E[\bm x \sim \Dirichlet]{\prod_{i = 1}^N x_i^{k_i}} = \frac{\Gamma(\Vert \bm\alpha\Vert_1)}{\Gamma(\Vert \bm\alpha\Vert_1 + \Vert \bm k\Vert_1)} \prod_{i = 1}^N \frac{\Gamma(\alpha_i + k_i)}{\Gamma(\alpha_i)}.
\end{equation}

In particular
\begin{align}
    \E[\bm x \sim \Dirichlet]{x_i} & = \frac{\alpha_i}{\Vert \bm\alpha\Vert_1} \\ 
    \Var[\bm x \sim \Dirichlet]{x_i} & = \frac{\alpha_i\left( 1 - \frac{\alpha_i}{\Vert \bm\alpha\Vert_1}\right)}{\Vert \bm\alpha\Vert_1 (\Vert \bm\alpha\Vert_1 + 1)} \\ 
    \Cov[\bm x \sim \Dirichlet]{x_i, x_j} & = \frac{-\alpha_i\alpha_j}{\Vert \bm\alpha\Vert_1^2 ( \Vert \bm\alpha\Vert_1 + 1)} 
\end{align}

\subsection{Proof of~\Cref{cor.observable}}\label{app.observable}
Let us express the wavefunction as the $\bm C$ matrix, such that the wavefunction is reconstructed from its elements as
\begin{equation}
    \ket{\psi(x)} = \sum_{j=1}^{2^n} \sum_{k=-K}^K c_{k,j} e^{ik\mu x} \ket{j},
\end{equation}
where $\ket j$ is expressed in this case the eigenbasis of the observable of interest $H$. We are interested in the function
\begin{equation}
    h(x) = \bra{\psi(x)} H \ket{\psi(x)}, 
\end{equation}
which in the eigenbasis of $H$ is
\begin{equation}
    h(x) = \sum_j \sum_{k} \sum_{l} \lambda_j c_{k,j} c^*_{l,j} e^{i \mu (k - l)\ x}.
\end{equation}
We give a bound now on the terms sharing the same frequencies
\begin{equation}
     \left\vert \sum_j \sum_{k - l = \omega} \lambda_j c_{k,j} c^*_{l,j} \right\vert^2 \leq 
     \norm{H}_\lambda^2\left\vert \sum_j \sum_{k - l = \omega} c_{k,j} c^*_{l,j} \right\vert^2 \leq 
     \norm{H}_\lambda^2\sum_{k - l = \omega}\left(\sum_j \left\vert  c_{k,j}\right\vert^2 \right) \left(\sum_j \left\vert  c_{l,j}\right\vert^2 \right),
\end{equation}
where we used the triangular inequality and the Cauchy-Schwarz inequality. The last term is related to~\Cref{th.conv_spectrum}. Each of these elements is drawn from the Dirichlet distribution imposed by the spectrum $\K_g^{*L}$. The aggregation property of Dirichlet distributions allows us to directly work with the spectrums. The spectrum of interest is a modified convolution of $\K_g^{*L}$ with itself under an inversion of the variable, namely
\begin{equation}
    \sum_{k - l = \omega} \K_g^{*L}(k) \K_g^{*L}(l) = \sum_k \K_g^{*L}(k) \K_g^{*L}(k - \omega) = \sum_k \K_g^{*L}(k) \K_g^{*L}(-(k - \omega)) = \left(\K_g^{*L} * \K_g^{\prime \ *L}\right) (\omega),
\end{equation}
with $\K_g^{*L}(x) =  \K_g^{\prime \ *L}(-x)$. In the case of symmetric spectra, both functions are equivalent. Recalling the properties of Dirichlet distributions, we can bound
\begin{equation}
      \norm{H}_\lambda^{-2} \left\vert a_\omega(\btheta) \right\vert^2\leq 
     p_\omega \sim \Dir(\K_g^{*2L}(\omega)),
\end{equation}
with 
\begin{equation}
    a_\omega(\btheta) = \sum_j \sum_{k - l = \omega} \lambda_j c_{k,j} c^*_{l,j}
\end{equation}
\qed

\subsection{Proof of~\Cref{th.av_lipschitz}}\label{app.av_lipschitz}
We use tools from statistics to compute upper and lower bounds to the Lipschitz constant of a hypothesis function $\Lambda(h_\btheta)$. 
We first recall the definition
\begin{equation}
    \Lambda(h_\btheta) := \sum_{k = -K}^K \mu \vert k \vert \vert a_k\vert, 
\end{equation}
and we recall the result from~\Cref{cor.observable} in the limit of many re-uploadings. We know that
\begin{equation}
    \E[\Btheta]{\Lambda(h_\btheta)} = \sum_{k = -K}^K \mu \vert k \vert \E[\Btheta]{\vert a_k \vert}.
\end{equation}

We can use the known bound for the probability distribution underlying $\vert a_k\vert$, $\vert a_k \vert^2 \leq p_k \sim \Dir(\K_g^{* L})$. In particular, the marginals of the Dirichlet distribution are beta distributions~\cite{bailey1992distributional}. The beta probability distribution with parameters $\alpha$ and $\beta$ is defined as
\begin{equation}
    \operatorname{Beta}_{\alpha, \beta}(x) = \frac{\Gamma(\alpha + \beta)}{\Gamma(\alpha) \Gamma(\beta)} x^{(\alpha - 1)} (1 - x)^{(\beta - 1)}.
\end{equation}
In our case, see~\Cref{eq.dirichlet}, $\alpha$ is given by the Gaussian spectrum and $\beta = 1$. The expectation value of each element is given by
\begin{equation}
    \E[\Btheta]{\vert a_k \vert} \leq \int_{0}^{1} dx \frac{\Gamma(\alpha_k + 1)}{\Gamma(\alpha_k) } x^{(\alpha_k - 1/2)} = \frac{\alpha_k}{\alpha_k + 1 / 2} \leq 2 \alpha_k,
\end{equation}
using the property of the gamma function $\Gamma(1 + x) = x \Gamma(x)$. The last inequality allows us to compute an upper bound in the limit of many re-uploadings by just computing
\begin{equation}
    \sum_{k = -K}^K \mu k \E[\Btheta]{\vert a_k \vert} \leq 2 \norm{H}_\lambda \sum_{k = -K}^K \mu k \K_g^{*2L}(k) \approx 4 \norm{H}_\lambda \int_{0}^\infty \frac{\mu k}{\sqrt{4\pi L} \sigma_g} \exp\left( -\frac{k^2}{4 \sigma^2_g L}\right) = \norm{H}_\lambda \frac{4}{\sqrt{\pi}} \sigma_g \mu \sqrt{L},
\end{equation}
leading to the first result of the theorem.

The lower bound is easy to obtain by recalling the property $\norm{a}_1 \geq \norm{a}_2$. In our context, and in the limit of Gaussian processes
\begin{equation}
    \E[\Btheta]{\Lambda(h_\btheta)}^2 \geq \norm{H}_\lambda^2 \sum_{k = -K}^K \mu^2 k^2 \E[\Btheta]{\vert a_k\vert^2}\approx \norm{H}_\lambda^2 2 L \mu^2 \sigma_g^2, 
\end{equation}

leading to the second result of the theorem: $\E[\Btheta]{\Lambda(h_\btheta)} \geq \norm{H}_\lambda \sigma_g \mu \sqrt{2 L}$.\qed

\subsection{Proof of~\Cref{th.dev_lipschitz}}\label{app.dev_lipschitz}
We are interested in knowing the probability of $\Lambda(h_\btheta)$ to be larger than a certain reference value by some distance. We take this reference value to be the average $\E[\Btheta]{\Lambda(h_\btheta)}$, as in many statistics results. Consider now the lower-bound on the expectation value from~\Cref{th.av_lipschitz}. Since
\begin{equation}
    \Lambda(h_\btheta) - \E[\Btheta]{\Lambda(h_\btheta)} \geq t \implies \Lambda(h_\btheta) - \norm{H}_\lambda \mu \sqrt{2 L}\sigma_g \geq t, 
\end{equation}
but not in the opposite direction, then 
\begin{equation}
    \operatorname{Prob}_\Btheta\left( \Lambda(h_\btheta) - \E[\Btheta]{\Lambda(h_\btheta)} \geq t\right)\leq \operatorname{Prob}_\Btheta\left(\Lambda(h_\btheta) - \norm{H}_\lambda \mu \sqrt{2 L}\sigma_g \geq t\right). 
\end{equation}
We can bound the right hand by considering Hoeffding's inequality~\cite{hoeffding1963probability}. Let ${X_i}$ be a set of independent random variables, and let $X_i \in [a_i, b_i]$ {\sl almost surely}, then
\begin{equation}
    \operatorname{Prob}\left( \sum_i X_i - \E{\sum_i X_i} \geq t \right) \leq \frac{1}{2}\exp\left(\frac{-t^2}{\sum_i(a_i - b_i)^2} \right).
\end{equation}

Hoeffding's inequality cannot be directly applied to a Dirichlet distribution since the variables are not independent. However, this problem can be overcome for this particular case by recalling the following property. If ${X_i} \sim \operatorname{Dir}(\alpha_i)$, then
\begin{equation}
    X_i \sim \frac{Y_i}{V}, 
\end{equation}
with
\begin{align}
    Y_i & \sim \operatorname{Gamma}(\alpha_i, \theta) \\
    V = \sum_i Y_i & \sim \operatorname{Gamma}\left(\sum_i \alpha_i,\theta\right)
\end{align}
By changing the description of the Dirichlet distribution to the quotient of gamma distributions we can now apply Hoeffding's inequality. Without loss of generality, we can assume that all probabilities are bounded between $0$ and $1$, thus we can find a first bound by recalling
\begin{equation}
    \sum_n n^2 \in \mathcal{O}((\norm{g}_\lambda L)^3), 
\end{equation}
and thus
\begin{equation}
    \operatorname{Prob}_\Btheta\left( \Lambda(h_\btheta) - \E[\Btheta]{\Lambda(h_\btheta)} \geq t\right) \leq \operatorname{Prob}_\Btheta\left(\Lambda(h_\btheta) - \norm{H}_\lambda \sqrt{2 L} \mu\sigma_g \geq t\right) \in \bigO{\exp\left(\frac{-t^2}{(\norm{g}_\lambda L)^3}\right)}
\end{equation}
\qed

\subsubsection{A tighter numerical bound}\label{app.num_bound}
\begin{figure}[b!]
    \centering
    \subfigure[~Numerical calculations for~\Cref{eq.cdf} for decreasing values of $\alpha$. The values of the random variable concentrate in small values for $t$ as $\alpha$ decays.]{\includegraphics[width=.45\linewidth]{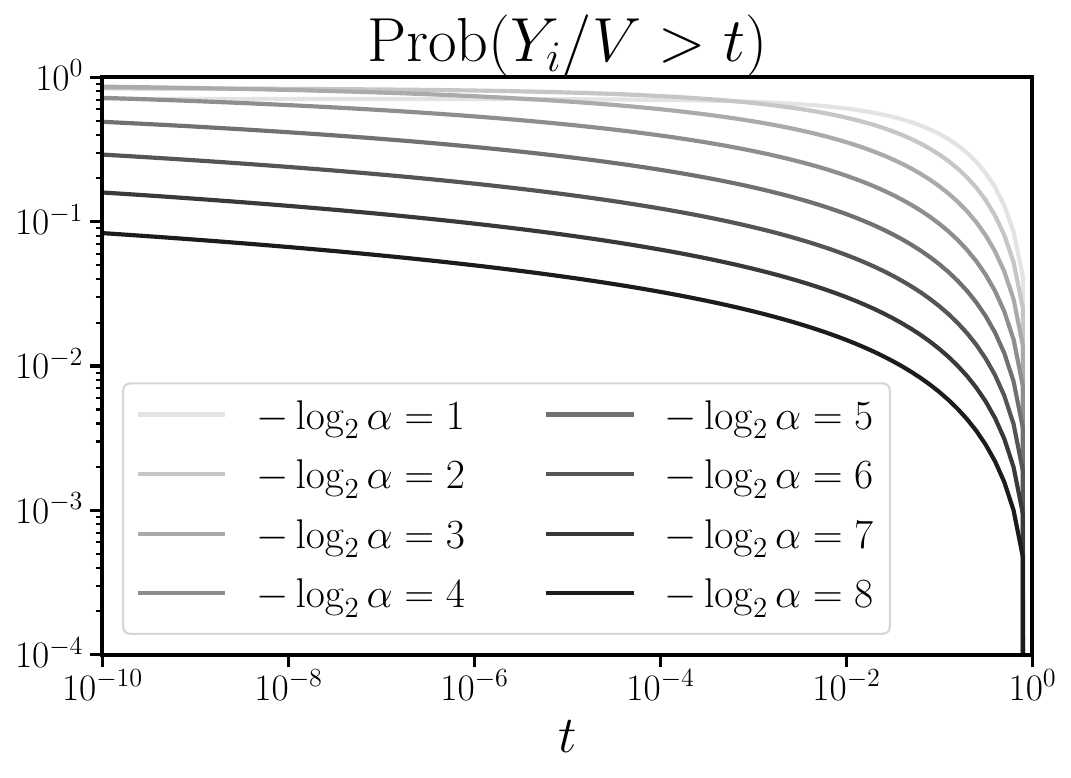}
    \label{fig.subgaussianity1}}
    \subfigure[~Numerical approximation to~\Cref{eq.sum_n_lipschitz} for increasing $\sigma_g \sqrt{L}$. The value $\epsilon$ is set in this calculations to $10^{-10}$]{\includegraphics[width=.45\linewidth]{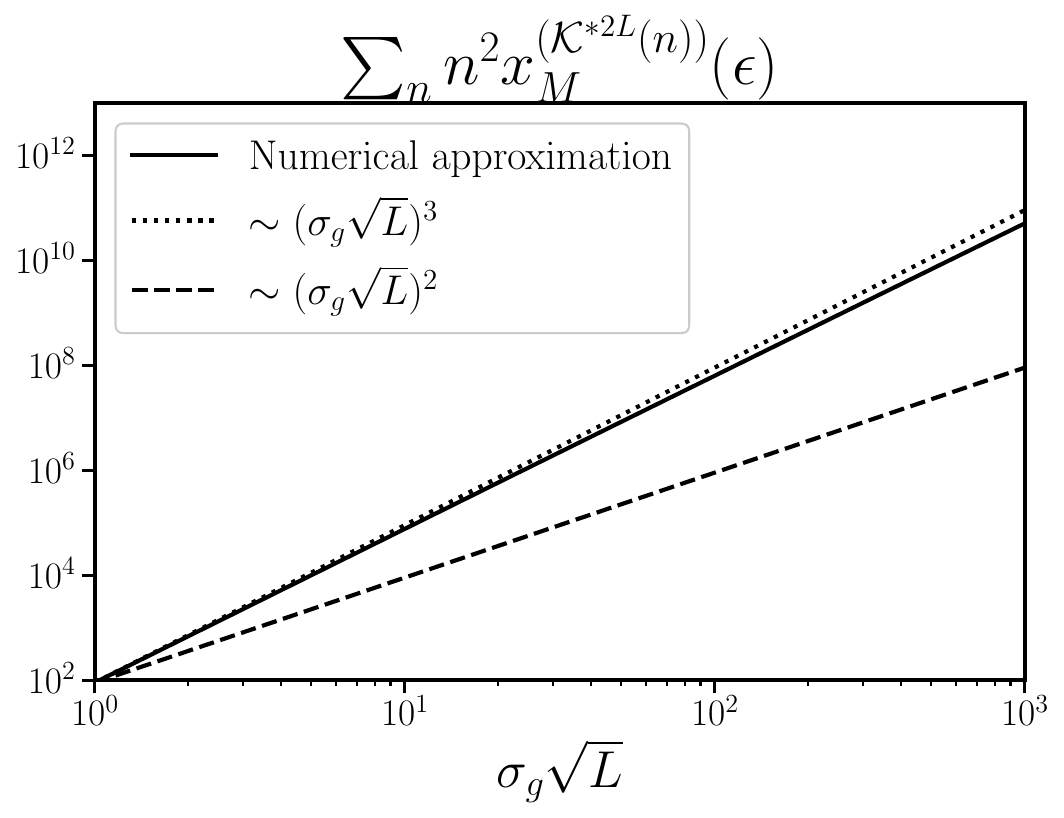}
    \label{fig.subgaussianity2}}
    \caption{Numerical auxiliary calculations for~\Cref{eq.cdf} and~\Cref{eq.sum_n_lipschitz}. These results substitute the non-accessible analytical treatment of the probability distributions of interest to obtain the bound given in~\Cref{eq.tight_bound_lipschitz}}
    \label{fig.sugbaussianity}
\end{figure}

This bound can be however easily improved by recalling subgaussianity properties of the Gamma distribution. A random variable $X$ is subgaussian if its cumulative distribution function decays faster than exponentially
\begin{equation}
    \operatorname{Prob}\left( \vert X \vert \geq t\right) \in \bigO{\exp\left( -t^2\right)}, 
\end{equation}
for some positive constant $K$. We can compute this cumulative probability for the quotient of Gamma distributions as
\begin{equation}\label{eq.cdf}
    \operatorname{Prob}\left( \frac{Y_i}{V} \geq t\right) = \int_0^\infty dx \int_x^{x/t} dy \frac{x^{\alpha - 1} e^{-x} e^{-y}}{\Gamma(\alpha)} = \frac{1}{2^{\alpha}} - \frac{1}{(1 + t^{-1})^{\alpha}} \in \bigO{\exp\left( -t^2\right)}.
\end{equation}
These functions take the value $1$ for $t=0$ and decay until vanishing for $t=1$. The decay is faster as $\alpha\rightarrow 0$, as it can be seen in~\Cref{fig.subgaussianity1}. We can thus recover Hoeffding's inequality with the observation that each $X_i$ is bounded by the function in~\Cref{eq.cdf}. In particular, the variable $X_i$ is, with probability $1 - \epsilon$, smaller than
\begin{equation}\label{eq.variable_bound}
    x^{(\alpha_i)}_M(\epsilon) = \left( \left( \frac{1}{2^{\alpha_i}} - \epsilon \right)^{-1 / \alpha_i} - 1\right)^{-1}.
\end{equation}
For a sufficiently small $\epsilon$, the denominator of the exponent of Hoeffding's inequality becomes 
\begin{equation}\label{eq.sum_n_lipschitz}
    \sum_n(a_n - b_n)^2 = 2 \sum_{n = 1}^{\norm{g}_2 L} n^2  x^{(\K_g^{* 2 L}(n))}_M(\epsilon), 
\end{equation}
with $\K_g^{*2L}$ a Gaussian spectrum in the limit of large $R$. The Gaussian limit forces the intuition that only a small number of elements will contribute effectively, while for large values of $n$ the corresponding Dirichlet variable is always so small that it has negligible influence in the Lipschitz constant. The description of the variable bounds in~\Cref{eq.variable_bound} and the sum in~\Cref{eq.sum_n_lipschitz} prevent a straightforward analysis in terms of the relevant quantity $\sigma_g \sqrt{R}$. We can however make a numerical analysis, depicted in~\Cref{fig.subgaussianity2}. This calculation shows that the sum in~\Cref{eq.sum_n_lipschitz} follows a polynomial trend in $\sigma_g \sqrt{R}$, which is the variance of the resulting Gaussian spectrum. Therefore, we can update our previous version of Hoeffding's inequality to
\begin{equation}
    \operatorname{Prob}_\Btheta\left(\Lambda(h_\btheta) - \norm{H}_\lambda \sqrt{2 L}\sigma_g \geq t\right) \in \bigO{\exp\left(\frac{-t^2}{(\sigma_g \sqrt{L})^3}\right)}.
\end{equation}

\subsection{Extension to non-harmonic spectrum}\label{app.nonharmonic}
The non-harmonic extension leads to an average of the elements in the trigonometric polynomial given by
\begin{equation}
    \E[\Btheta]{\vert a_{\vec k} \vert ^2 } = \frac{1}{\sqrt{(4\pi L)^D \vert \Sigma \vert}} \exp\left( -\frac{\vec k^T \Sigma^{-1} \vec k}{4 L}\right),
\end{equation}
where $\vert \Sigma \vert$ is the determinant of $\Sigma$.
We compute now $\Lambda(h_\btheta)$ following the steps from~\Cref{app.av_lipschitz}, given by
\begin{equation}
    \E[\Btheta]{\Lambda(h_\btheta)} = \sum_{\vec k} \vert \vec \mu \cdot \vec k \vert  \E[\Btheta]{\vert a_{\vec k}\vert}  \leq \frac{2}{\sqrt{(4\pi L)^D \vert \Sigma \vert}} \int_{\mathbb R^D} {d\vec k}\vert \vec \mu \cdot \vec k \vert \exp\left( -\frac{\vec k^T \Sigma^{-1} \vec k}{4 L}\right) .
\end{equation}
Notice that $d\vec k$ integrates over $D$-dimensional space. 
We perform now a change the variables to diagonalize $\Sigma = U^{\dagger} S U$, and consequently choose $\vec l=U \vec k$. The diagonal elements of $S$ are denoted $\{s_j^2\}_j$ The quantity of interest is now $\vec\mu \cdot (U^\dagger \vec l) = (U \vec\mu) \cdot \vec l$. Since $U$ is unitary ${d\vec l} = {d\vec k}$
\begin{align}
    \E[\Btheta]{\Lambda(h_\btheta)} & \leq \frac{2}{\sqrt{(4\pi L)^D \vert \Sigma \vert}}\int_{\mathbb R^D} d\vec l \vert (U\vec \mu) \cdot \vec l \vert \exp\left( -\frac{\vec l^T S^{-1} \vec l}{4 L}\right) \leq \\
    & \frac{2}{\sqrt{(4\pi L)^D \vert \Sigma \vert}} \sum_{j = 1}^D  \vert (U \vec \mu)_j \vert \int_{\mathbb R^D} d\vec l \vert l_j \vert \exp\left( -\frac{\vec l^T S^{-1} \vec l}{4 L}\right) \label{eq.integral_multidim}
\end{align}
We focus now on the integral. 
\begin{align}
    \int_{\mathbb R^D} d \vec l \vert l_j \vert \exp\left( -\frac{\vec l^T S^{-1} \vec l}{4 L}\right) = & \int_{\mathbb R} dl_j \vert l_j \vert \exp\left( -\frac{\vec l_j^2}{4 L s_j^2}\right)\prod_{i\neq j} \int_{\mathbb R} d l_i \exp\left( -\frac{\vec l_i ^2}{4 L s_i^2}\right)\\
    & = 4 L s_j^2  \prod_{i\neq j} \sqrt{4 \pi L s_i^2} = 
    \frac{1}{\pi}\sqrt{(4\pi L)^{D+1}} \sqrt{\vert \Sigma \vert} s_j
\end{align}
Plugging this result into~\Cref{eq.integral_multidim} we obtain
\begin{equation}
    \E[\Btheta]{\Lambda(h_\btheta)} \leq \frac{(2\sqrt{L \pi})^{D+1} }{\pi} \frac{\sqrt{\vert \Sigma \vert} \vert \vec \mu  U^{\dagger} \vert \cdot \sqrt{\vec S}}{\sqrt{(4\pi L)^D \vert \Sigma \vert}} = \frac{4\sqrt{L}}{\sqrt{\pi}} \sum_{j=1}^D \vert U \vec \mu\vert_j s_j
\end{equation}
By means of Cauchy-Schwarz inequality, we can give a looser yet more comprehensive bound as
\begin{equation}
    \E[\Btheta]{\Lambda(h_\btheta)}  \leq \frac{4}{\sqrt \pi} \norm{\vec \mu}_2\sqrt{\Tr(\Sigma)} \sqrt{L}. 
\end{equation}

For the lower bound we follow~\Cref{app.av_lipschitz} to obtain 
\begin{equation}
    \E[\Btheta]{\Lambda(h_\btheta)}^2 \geq \norm{H}_\lambda^2 \sum_{k = -K}^K (\vec \mu \cdot \vec k)^2 \E[\Btheta]{\vert a_k\vert^2}. 
\end{equation}
We recall the property~\cite{zee2010quantum, user268722012answer}
\begin{equation}
\int d^D \vec k f(\vec k) \exp\left(-\frac{\vec k^T \Sigma^{-1} k}{2} \right) = \sqrt{(2\pi)^D \vert \Sigma \vert}\left.\exp\left(\frac{\vec \nabla^T \Sigma^{-1} \vec\nabla}{2} \right) f(\vec k)\right\vert_{\vec k = 0},
\end{equation}
where $\vec \nabla_j = \partial / \partial \vec k_j$. Since $f(\vec k) = (\vec \mu \cdot \vec k)^2$, we can reduce
\begin{equation}
    \left.\exp\left(\frac{\vec \nabla^T \Sigma^{-1} \vec\nabla}{2} \right) f(\vec k)\right\vert_{\vec k = 0} = \vec \mu^T \Sigma \vec \mu \geq \norm{\vec\mu}_2^2 \min_\Lambda(\Sigma), 
\end{equation}
yielding a result
\begin{equation}
    \Lambda^2(h_\btheta) \geq \norm{H}_\lambda^2 2 L \norm{\mu}_2^2 \min_\lambda(\Sigma).
\end{equation}
\qed

\subsubsection{A simple example}

We illustrate the spectral convolution with an example. Consider a data generator whose spectrum and multiplicities are 
\begin{align}
    \lambda & = \{-\sqrt{2}-1 , -\sqrt{2} ,-1, 0, +1 ,+\sqrt{2},+\sqrt{2}+1\} \\
    m(\lambda) & = \{1,1,1,2,1,1,1\}
\end{align}

Any frequency resulting from the $L$-fold application of such data generator can be written as $\lambda_{k,l} = k \sqrt{2} + l$ where $-L\leq k,l \leq L$ are integers. The corresponding frequency content can therefore be represented as a two-dimensional tensor $A$. The elements of $A$ follow a 2-dimensional Dirichlet distribution, in the sense of \Cref{th.conv_spectrum}, given by the convoluted kernel

\begin{equation}
\K_g^{* L}  = 
    \frac{1}{8}\begin{bmatrix}
    1 & 1 & 0\\
    1 & 2 & 1\\
    0 & 1 & 1
    \end{bmatrix} ^{*L}
\end{equation}

In the limit of large $L$, the central limit theorem applies exactly in the same way as in the harmonic case, and the $L$-fold convolution tends towards a multivariate Gaussian kernel with $[0,0]$ mean and covariance matrix
\begin{equation}
    \Sigma = \frac{L}{2} \begin{pmatrix}
        1  & 0.5 \\ 0.5 & 1 
    \end{pmatrix}.
\end{equation}

\end{document}